\documentclass[useAMS,usenatbib]{mn2e}
\usepackage{epsfig}
\usepackage{overpic}

\newcommand{\cii}{[C\,{\sc ii}] }
\newcommand{\ciino}{[C\,{\sc ii}]}
\newcommand{\oi}{[O\,{\sc i}] }
\newcommand{\oino}{[O\,{\sc i}]}

\title[CO-dark gas -- II: The temperature of the gas]{CO-dark gas and molecular filaments in Milky Way-type galaxies -- II: The temperature distribution of the gas}
\author[Glover \& Smith]{Simon C. O. Glover$^{1}$ \& Rowan J. Smith$^{2}$ \\
$^{1}$Universit\"{a}t Heidelberg, Zentrum f\"{u}r Astronomie, Institut f\"{u}r Theoretische Astrophysik, Albert-Ueberle-Stra{\ss}e 2, \\
\mbox{ } 69120 Heidelberg, Germany \\
$^{2}$Jodrell Bank Centre for Astrophysics, School of Physics and Astronomy, University of Manchester, Oxford Road, \\
\mbox{ } Manchester, M13 9PL, UK \\
}

\begin{document}
\maketitle

\begin{abstract}
We investigate the temperature distribution of CO-dark molecular hydrogen (H$_{2}$) in 
a series of disk galaxies simulated using the {\sc Arepo} moving-mesh code. In conditions
similar to those in the Milky Way, we find that H$_{2}$ has a flat temperature distribution 
ranging from 10--100~K. At $T < 30$~K, the gas is almost fully molecular and has a high 
CO content, whereas at $T > 30$~K, the H$_{2}$ fraction spans a broader
range and the CO content is small, allowing us to classify gas in these two regimes as 
CO-bright and CO-dark, respectively. The mean sound speed in the CO-dark H$_{2}$ 
is $c_{\rm s, dark} = 0.64 \: {\rm km \: s^{-1}}$, significantly lower than the value 
in the cold atomic gas ($c_{\rm s, cnm} = 1.15 \: {\rm km \: s^{-1}}$), implying that
the CO-dark molecular phase is more susceptible to turbulent compression and gravitational
collapse than its atomic counterpart. We further show that the temperature of the CO-dark H$_{2}$ 
is highly sensitive to the strength of the interstellar radiation field, 
but that conditions in the CO-bright H$_{2}$ remain largely unchanged. Finally, we examine
the usefulness of the \cii and \oi fine structure lines as tracers of the CO-dark gas. We show 
that in Milky Way-like conditions, diffuse \cii emission from this gas should be detectable. 
However, it is a problematic tracer of this gas, as there is only
a weak correlation between the brightness of the emission and the H$_{2}$ surface density.
The situation is even worse for the \oi line, which shows no correlation with the H$_2$ surface
density.
\end{abstract}


\section{Introduction}
Star formation within our Galaxy appears to occur exclusively within clouds of molecular gas. It is therefore important to understand the physical conditions within these clouds if we want to understand the role they play in star formation. Ideally, we would do this by studying the line emission from molecular hydrogen (H$_{2}$), which is by far the most abundant molecule within these clouds. Unfortunately, however, the temperature within the bulk of these clouds is too low to excite even the lowest accessible rotational transition of H$_{2}$, meaning that H$_{2}$ emission is useful only for tracing atypical regions such as the shock fronts created by protostellar outflows \citep[see e.g.][]{rb01}. Consequently, if we want to learn about the physical conditions within the clouds (temperatures, densities, velocity structure, etc.), we have to rely on information provided by other observational tracers.

The most widely used tracer molecule is carbon monoxide (CO). This is the second most abundant molecular species within typical molecular clouds and its lowest rotational transitions are easily excited at the gas temperatures found there. Observations of CO line emission have therefore become one of the main ways in which we study the properties of both Galactic and extragalactic molecular clouds \citep[see e.g.][]{jackson06,hughes13}. However, chemical modelling of clouds has long made clear that CO is not a perfect tracer of H$_{2}$ \citep[see e.g.][]{th85,vdb88}. H$_{2}$ is far more effective than CO at shielding itself against the effects of ultraviolet photodissociation, and as a result the transition from atomic to molecular hydrogen within a cloud illuminated by the interstellar radiation field (ISRF) does not necessarily occur in the same location as the transition from C$^{+}$ to C to CO. Consequently, observations of CO alone miss a significant fraction of the H$_{2}$.

This CO-poor or ``CO-dark'' molecular gas has attracted increasing interest in the past few years, prompted by the realization that it may represent a significant fraction of the total molecular gas budget of the Milky Way. Observations \citep{grenier05,allen12,paradis12,langer14} and theoretical studies \citep{wolf10,smith14} suggest that in conditions typical of the local ISM, anywhere between 30\% and 70\% of the total mass of H$_{2}$ may be located in regions that are CO-poor and that are hence not traced by observations of CO emission.

Observational constraints on the temperature distribution of the molecular gas in the ISM primarily come from observations of molecules such as CO itself or ammonia (NH$_{3}$) that are only found in large quantities within CO-rich gas. These observations therefore only constrain the temperature of the CO-rich molecular gas. The temperature distribution in the CO-dark phase is much harder to constrain observationally. In cases where there is a bright ultraviolet (UV) background source (e.g.\ an AGN), H$_{2}$ can be observed in absorption via its Lyman-Werner band transitions, allowing one to constrain its temperature, but these observations only probe gas in a very small fraction of the ISM \citep[see e.g.][]{rach09}.
Since the CO-dark gas is much less well-shielded against the effects of the ISRF than the CO-rich gas, there are good reasons for believing that its temperature distribution could be significantly different from that in the CO-rich regions, an assumption which seems to be borne out by the high ($T \sim 50$--100~K) H$_{2}$ temperatures recovered from UV absorption line studies. If the temperature of the CO-dark gas does significantly differ from that of the CO-rich gas, then this is important as the temperature of the gas controls its sound-speed, and hence influences its stability against gravitational collapse as well as its response to the effects of supersonic turbulence \citep{pnj97,molina12}. The temperature of the molecular gas also affects the rate at which chemical reactions take place within it and so is an important input into any chemical model of the CO-dark molecular phase. Finally, the temperature of the CO-dark molecular gas also strongly influences the ease with which it can be detected using the atomic fine structure transitions of ionized carbon (C$^{+}$) and atomic oxygen (O).

In an effort to better understand the properties of the CO-dark molecular gas in galaxies like our own Milky Way, we have carried out detailed hydrodynamical simulations of a representative portion of the disk of a massive spiral galaxy \citep{smith14}. These simulations combine a state-of-the-art hydrodynamical code with a detailed treatment of the small-scale cooling physics and chemistry, and allow us to study the properties of the molecular phase in unprecedented detail. In this paper, we study in detail the temperature distribution of the molecular gas in these simulations, and assess its detectability using the fine structure lines of C$^{+}$ and O. 

The structure of our paper is as follows. In Section~\ref{sims}, we briefly summarize the numerical method used for our simulations and the initial conditions that we adopt. (A more lengthy discussion of both can be found in \citealt{smith14}). In Section~\ref{sec:MW},  we investigate the temperature distribution of the H$_{2}$ in our fiducial ``Milky Way'' simulation, and show that the temperatures that we derive for the CO-dark molecular component are in good agreement with previous theoretical predictions.
In Section~\ref{sec:change}, we extend the scope of our study to our other three simulations, allowing us to assess the effect of changing the mean gas surface density and the strength of the interstellar radiation field. In Section~\ref{sec:map}, we use a simple approximation to estimate the surface brightness of our simulated galaxies in the \cii 158$\,\mu$m, \oi 63$\,\mu$m and \oi 145$\,\mu$m fine structure lines and investigate how well the emission in these lines traces the CO-dark molecular gas. We discuss our results in Section~\ref{sec:discuss} (including a discussion of some possible caveats) and present our conclusions in Section~\ref{sec:conc}.

\section{Simulations}
\label{sims}
\subsection{Numerical approach}
The galactic disk simulations analyzed in this paper are the same as those presented in \citet{smith14}, and so we give only a brief description here and refer interested readers to that paper for more details. The simulations were carried out using a modified version of the {\sc Arepo} moving-mesh code \citep{springel10}. {\sc Arepo} is a novel hydrodynamical code that solves the fluid equations on an unstructured mesh, defined as the Voronoi tessellation of a set of mesh-generating points that can move freely with the gas flow. It combines many of the strengths of modern Eulerian hydrodynamical codes with those of smoothed particle hydrodynamics, and has been successfully applied to problems ranging from the formation of the first stars \citep{greif11,greif12} to the dynamical evolution of clouds close to the Galactic Centre \citep{bertram15}.

We have modified {\sc Arepo} to add a simplified treatment of the chemistry of the ISM, as discussed in detail in \citet{smith14}. We model the chemistry of hydrogen using the same treatment as in \citet{gm07a,gm07b}, and model the formation and destruction of CO using the approximate model introduced by \citet{nl97}. The radiative heating and cooling of the gas is treated using the detailed atomic and molecular cooling function introduced in \citet{glover10} and updated in \citet{gc12a}.

To model the attenuation of the interstellar radiation field due to H$_{2}$ self-shielding, CO self-shielding, the shielding of CO by H$_{2}$, and by dust absorption, we use the {\sc treecol} algorithm \citep{clark12}. This algorithm allows us to calculate a $4\pi$ steradian map of the dust extinction and the column densities of H$_{2}$ and CO surrounding each {\sc arepo} cell in an approximate but computationally efficient fashion. The resulting maps consist of 48 equal-area pixels, constructed using the {\sc healpix} algorithm \citep{healpix}. The functions to convert from H$_{2}$ and CO column densities to the corresponding shielding factors were taken from \citet{Draine96} and \citet{Lee96}, respectively. When constructing our column density and extinction maps, we include contributions only from gas at distances $L \leq 30$~pc from the cell of interest. This value is motivated by the fact that in the local ISM, the typical distance to the nearest UV-bright star is approximately  30~pc \citep{Reed00,ma01}.

We neglect the effects of self-gravity and stellar feedback. The lack of stellar feedback means that we will tend to over-produce molecular clouds, although this is offset to some extent by the absence of self-gravity, which will decrease the number of large clouds formed \citep[c.f.][]{girichidis15}. Note that absence of self-gravity is unlikely to be a major obstacle in forming H$_{2}$, as the transition from H to H$_{2}$ typically occurs at number densities $n < 100 \: {\rm cm^{-3}}$, where the contribution from self-gravity remains small in comparison to the effects of the global potential.

\subsection{Initial conditions}
The initial setup of each simulation is a torus of thickness 200~pc, inner radius 5~kpc, and outer radius 10~kpc. We do not include the inner portion of the disc, both for reasons of computational efficiency, and also because the behaviour of gas in this region will be strongly influenced by interactions with the galactic bar, if one is present. An analytical gravitational potential is applied to the torus to represent the influence of the stellar component of the galaxy. This potential is taken from \citet{Dobbs06} and combines a logarithmic potential producing a flat rotation curve with $v_{\rm rot} = 220 \: {\rm km \: s^{-1}}$ \citep{Binney87}, a potential for the outer halo taken from \citet{Caldwell81}, and a four-armed spiral component from \citet{Cox02}. The gas is initially in rotation at $v = v_{\rm rot}$ and develops pronounced spiral structure as it passes through the spiral potential (see Figure~\ref{fig:maps}).

As described in \citet{smith14}, we perform four simulations with different gas surface densities and radiation field strengths (see Table~\ref{tab:sims}). Our fiducial ``Milky Way'' model has an initial gas surface density $\Sigma = 10 \: {\rm M_{\odot}} \: {\rm pc^{-2}}$ and a radiation field strength given by \citet{draine78}. In addition, we explore the effects of decreasing the surface density while either keeping the radiation field strength fixed (the Low Density model) or also decreasing it (the Low \& Weak model). Finally, we also ran one simulation in which $\Sigma = 10 \: {\rm M_{\odot}} \: {\rm pc^{-2}}$ but the radiation field strength was increased by a factor of ten (the Strong Field model). We note that our assumption that the unattenuated radiation field strength is constant in both space and time is at best a crude approximation to the behaviour of the real ISM. However, \citet{parr03} have shown that away from regions of massive star formation, the variations in the radiation field strength in representative patches of the diffuse ISM are relatively small, around a factor of a few. Therefore our ``Milky Way'' and ``Strong Field'' simulations should bracket the behaviour of the real Milky Way ISM for the range of Galactocentric radii modelled in our simulations.

In all of our simulations, the gas was taken to have solar metallicity, with a carbon abundance\footnote{Unless otherwise noted, all fractional abundances quoted in this paper are measured with respect to the total number of hydrogen nuclei.} $x_{\rm C, tot} = 1.4 \times 10^{-4}$ and an oxygen abundance $x_{\rm O, tot} = 3.2 \times 10^{-4}$ \citep{sembach00}.  We assumed that the hydrogen and oxygen were initially in neutral atomic form, and that the carbon was present in the form of C$^{+}$, as in unshielded gas it is easily ionized by the ISRF. The cosmic ray ionization rate of atomic hydrogen was set to $\zeta_{\rm H} =  3 \times 10^{-17} \: {\rm s}^{-1}$ in all four simulations \citep{vv2000}.

To achieve the resolution necessary for accurately following the formation of H$_{2}$ and CO in the ISM, we make use of the system of mass refinement present in {\sc Arepo}. We focus our attention on a section of the disk, and simulate this section with an {\sc Arepo} cell mass of $4 \: {\rm M_{\odot}}$. In most of the remainder of the disk, we adopt a much larger cell mass of $1000 \: {\rm M_{\odot}}$ for reasons of computational efficiency. Between the high and low resolution regions, we have buffer zones in which we smoothly change the cell mass between $4 \: {\rm M_{\odot}}$ and $1000 \: {\rm M_{\odot}}$. Because we keep the cell mass constant, our spatial resolution varies as a function of the local gas density. At the mean density of a typical GMC, it is significantly less than a parsec in our high resolution region. This high spatial resolution allows us to begin to resolve the dense filaments and cores that contain most of the CO, and also means that the dense molecular clouds that form in our simulations are generally resolved with thousands or tens of thousands of  {\sc Arepo} cells. 

\begin{table}
        \centering
	\caption{Simulation parameters \label{tab:sims}}
                \begin{tabular}{l c c }
                 \hline
                 \hline
                  Simulation & Surface Density & Radiation Field Strength \\
                   & (M$_{\odot}$~pc$^{-2}$) & (relative to \citealt{draine78}) \\
                 \hline
                 Milky Way  & 10  & 1 \\
                 Low Density  & 4 & 1 \\
                 Strong Field & 10 & 10 \\
                 Low \& Weak & 4 & 0.1 \\
                 \hline
                 \hline
                \end{tabular}
\end{table}

\section{Molecular gas in the Milky Way simulation}
\label{sec:MW}
\subsection{Temperature distribution as a function of chemical composition}
\label{sec:tempd}
\begin{figure*}
\includegraphics[width=3.45in]{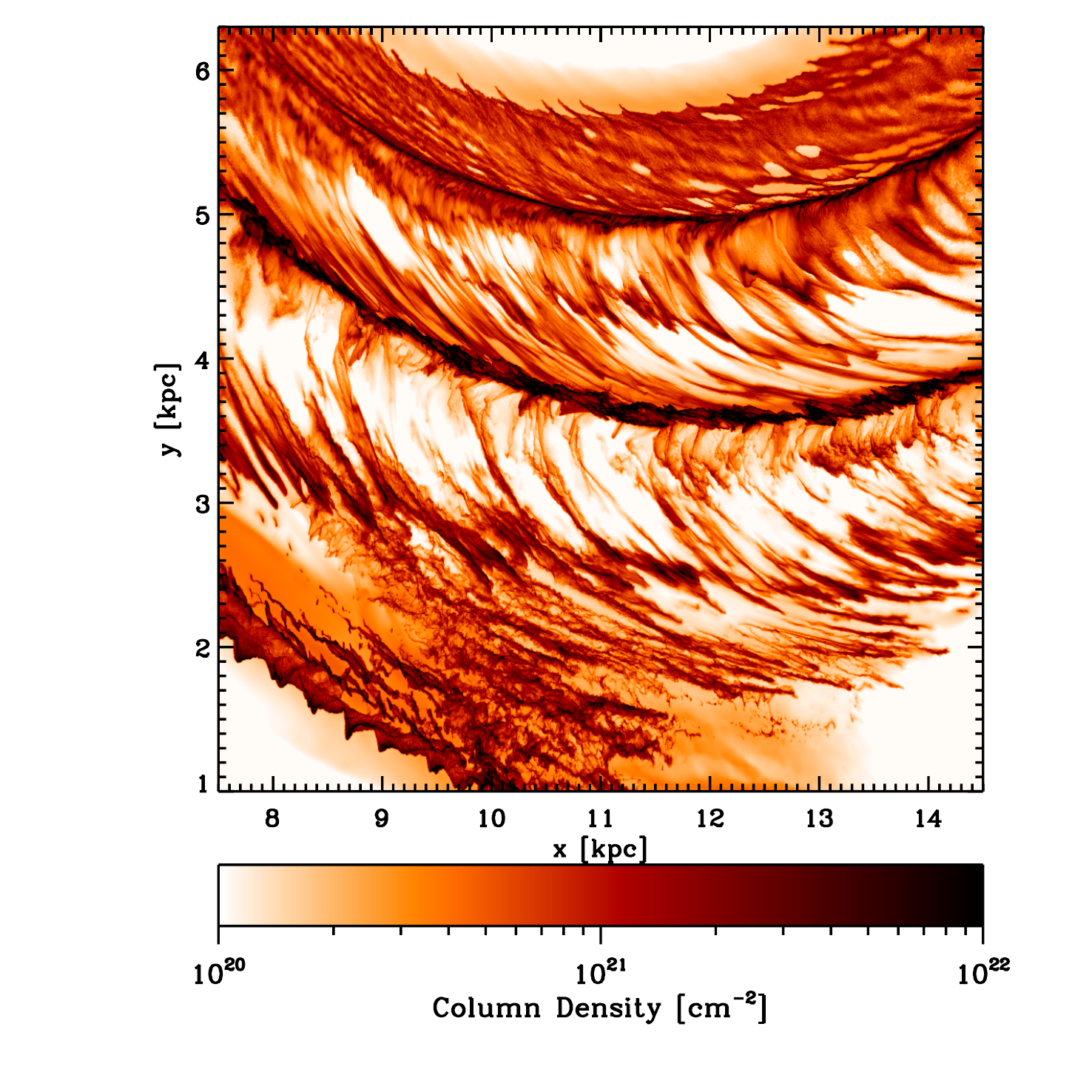}
\includegraphics[width=3.45in]{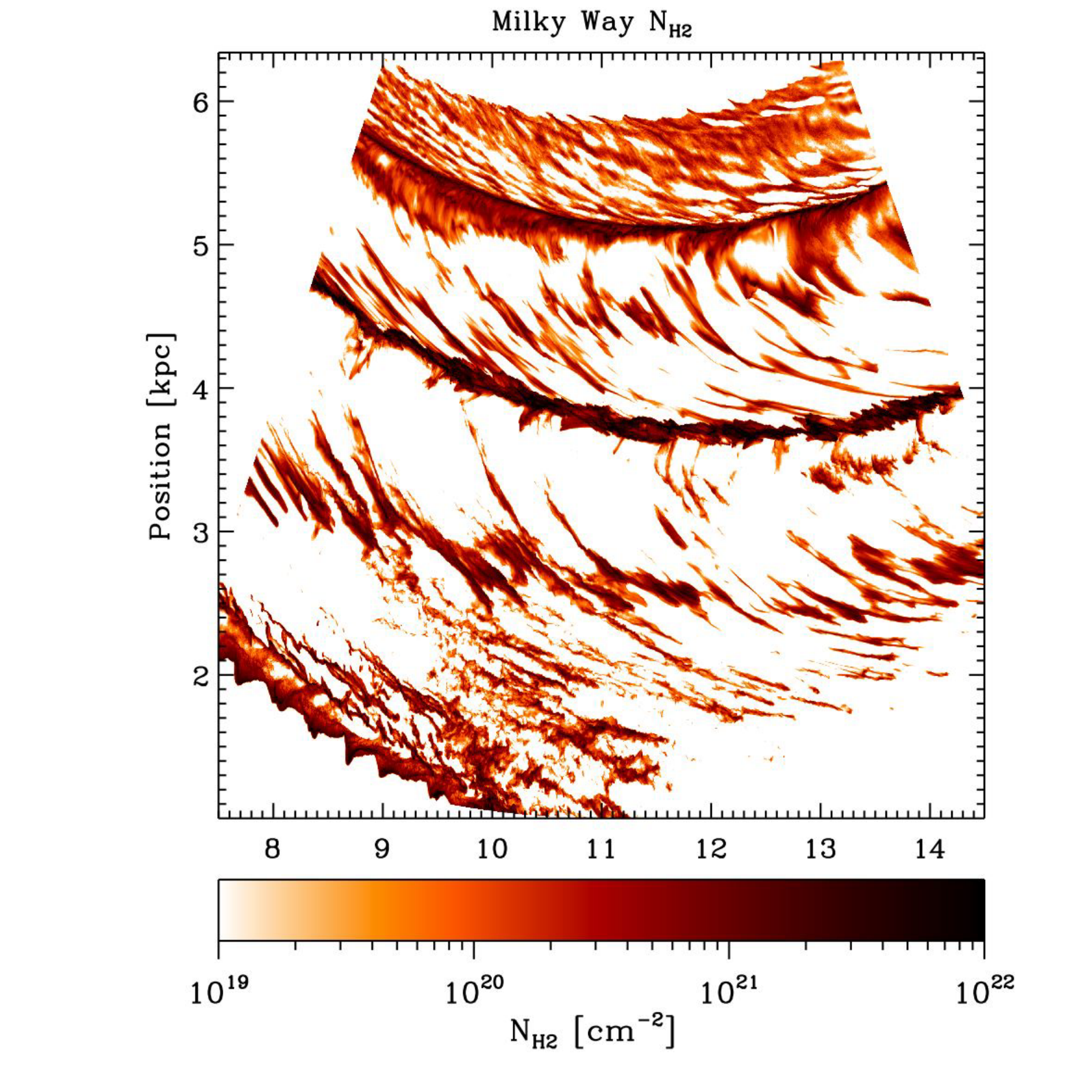}
\caption{Left: map of total column density of hydrogen nuclei in a region centered on the highly-resolved  section of the disc in the Milky Way simulation. 
Right: map of H$_{2}$ column densities for the same simulation. In this panel, we only show the gas within the highly-resolved region, as we do not
properly resolve H$_{2}$ formation in the remainder of the disc. In both panels, the coordinate axes show the $x$ and $y$ positions of the gas within the
simulation volume, with respect to an origin which we arbitrarily take to be at the bottom left of the volume.
 \label{fig:maps}}
\end{figure*}

We begin our study by examining the temperature distribution of the gas in our Milky Way simulation at a time $t = 261.1 \: {\rm Myr}$ after the beginning of the simulation. This time corresponds to one and a half complete rotations of our gas torus, which means that the gas has undergone six spiral arm passages. By this stage in the simulation, it has developed a considerable amount of dense, filamentary substructure, as illustrated in Figure~\ref{fig:maps}. Moreover, since this time is orders of magnitude longer than the cooling time in the dense ISM, we can be confident that most of the gas has reached thermal equilibrium.

\begin{figure*}
\begin{overpic}[width=3.2in]{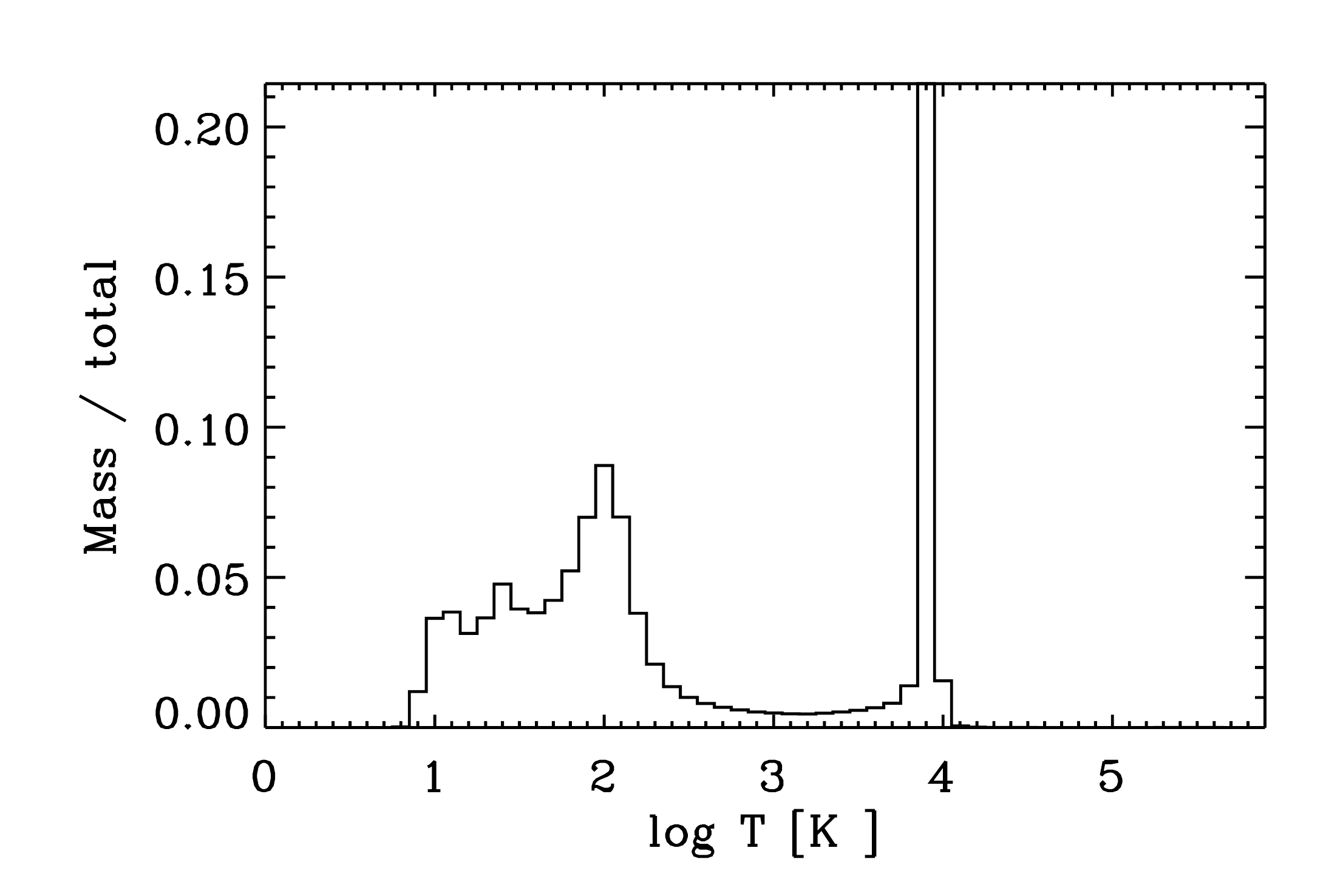}
\put(85,52){(a)}
\end{overpic}
\begin{overpic}[width=3.2in]{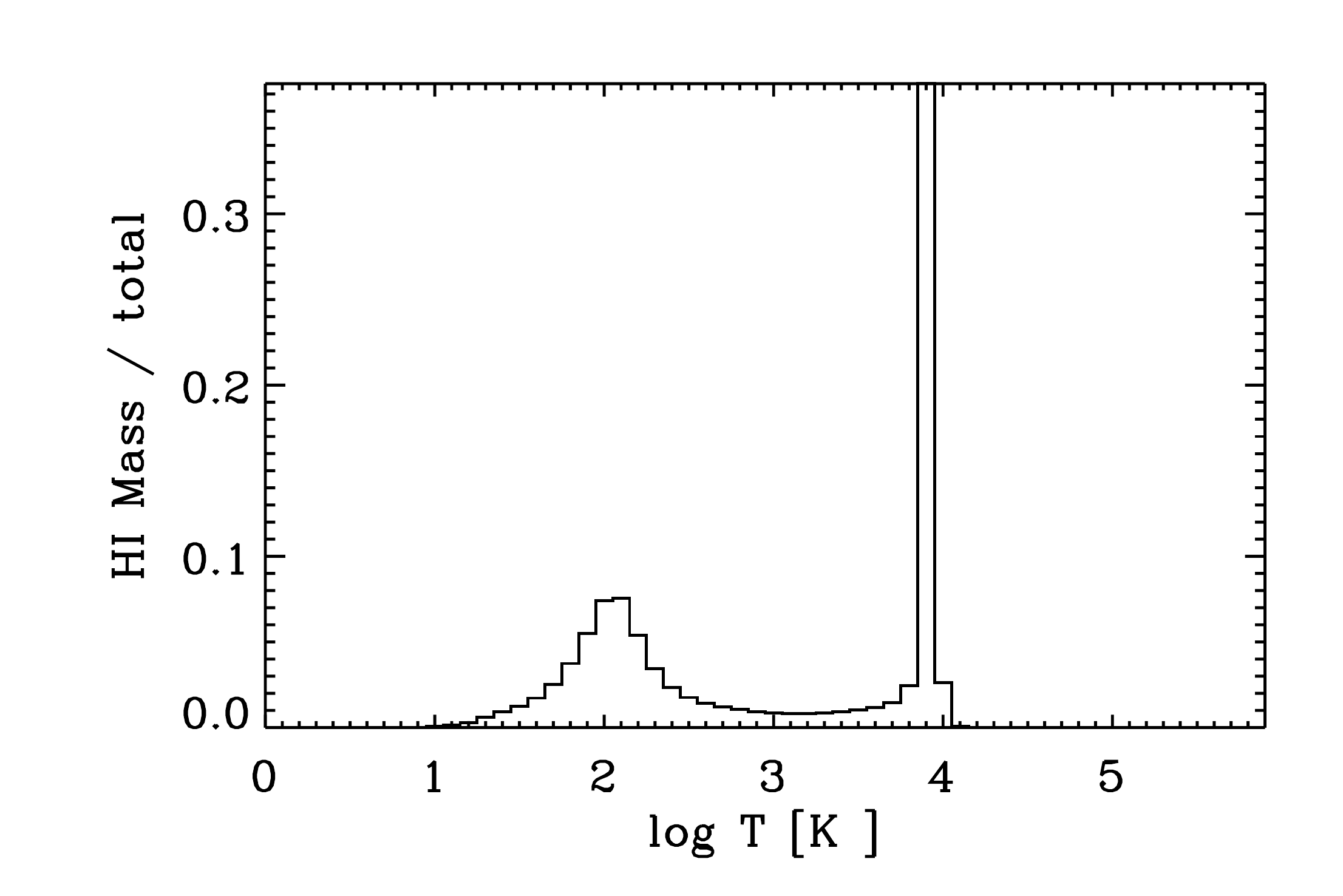}
\put(85,52){(b)}
\end{overpic}
\begin{overpic}[width=3.2in]{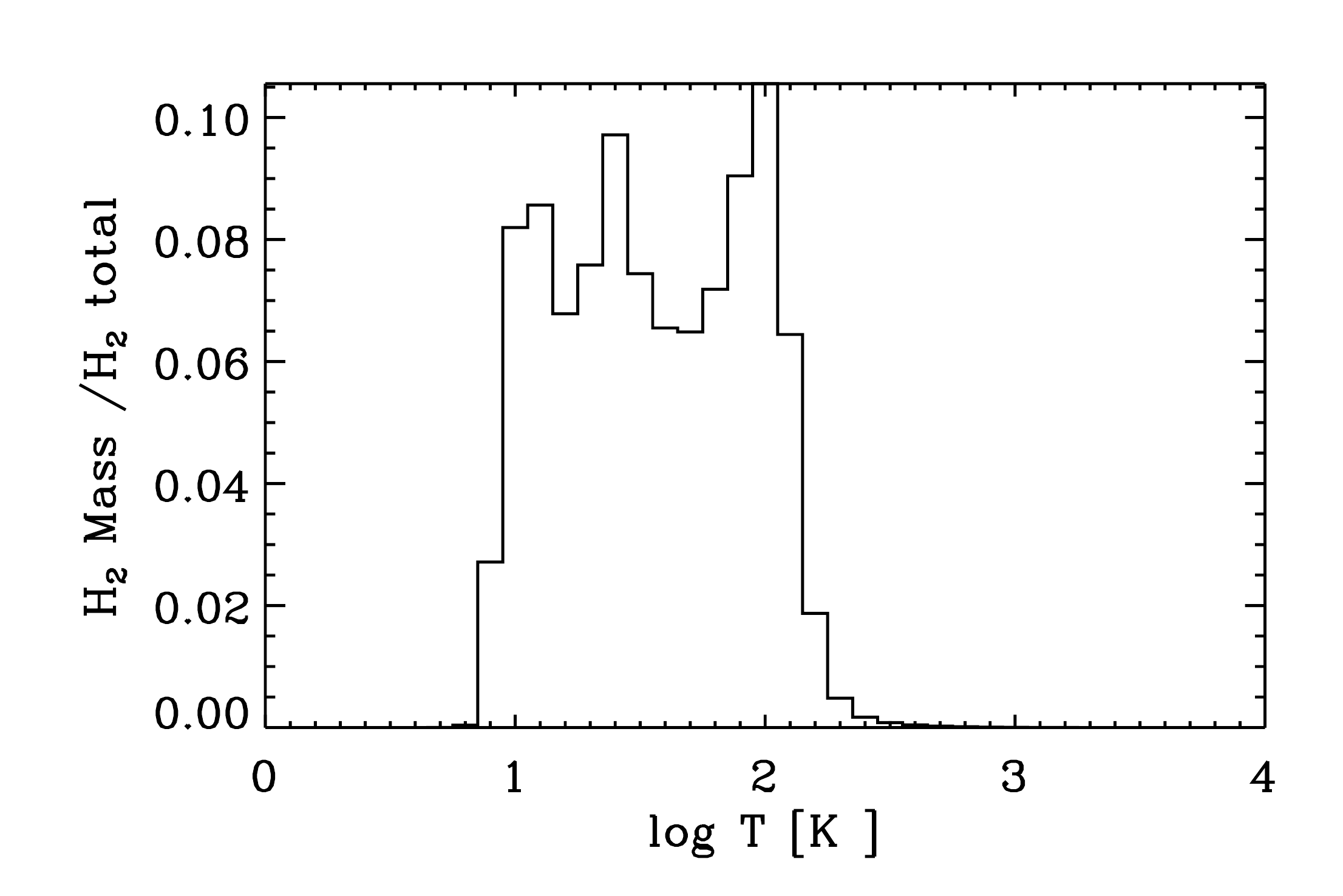}
\put(85,52){(c)}
\end{overpic}
\begin{overpic}[width=3.2in]{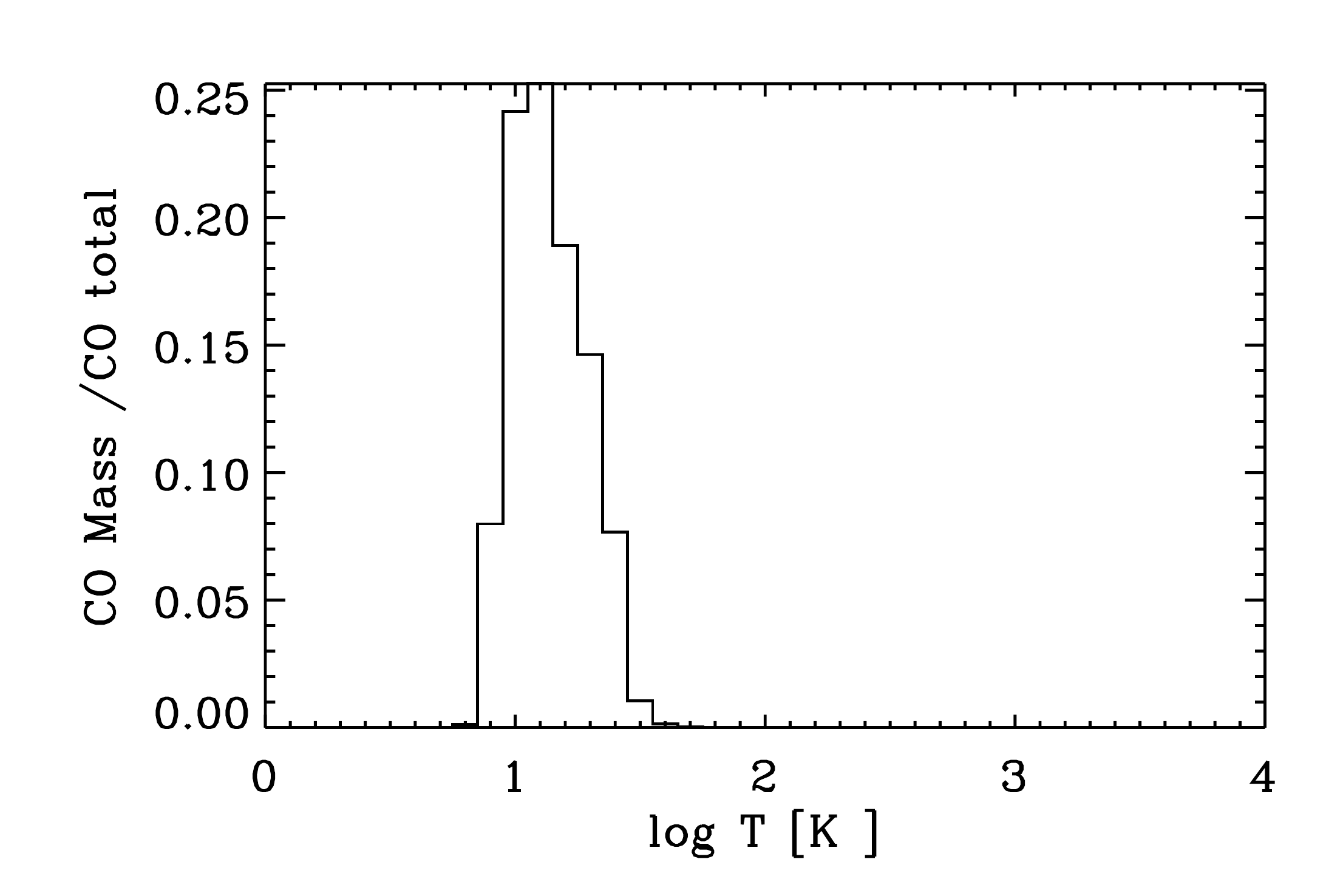}
\put(85,52){(d)}
\end{overpic}
\caption{(a) Mass-weighted histogram of the gas temperature in our Milky Way simulation at time $t = 261.1 \: {\rm Myr}$. The sharp peak just below $10^{4}$~K corresponds to the temperature at which Lyman-$\alpha$ cooling becomes ineffective. The absence of gas at higher temperatures is a consequence of the fact that our simulations do not include the effects of stellar feedback.  
(b) As (a), but weighted by the mass of atomic hydrogen. 
(c) As (a), but weighted by the mass of molecular hydrogen.
(d) As (a), but weighted by the mass of CO.
\label{fig:hist}}
\end{figure*}

Figure~\ref{fig:hist}a shows a mass-weighted histogram of the temperatures that we find in the gas. We see that the distribution is strongly bimodal: there is a pronounced peak at $T \sim 8000$~K, corresponding to the temperature below which Lyman-$\alpha$ cooling becomes ineffective and a second, more extended peak around 100~K. The high temperature peak is dominated by atomic gas (see Figure~\ref{fig:hist}b) and is easily identifiable as the warm neutral medium (WNM). The low temperature peak, on the other hand, consists of a mix of atomic gas (the cold neutral medium or CNM) and molecular gas. The absence of a hot phase with $T \gg 10^{4}$~K is a consequence of our neglect of stellar feedback in these simulations. The absence of feedback also means that there is less mixing of material between the cold and warm phases, and hence less gas at intermediate temperatures than found in simulations that include supernova feedback \citep[c.f.][]{hill12,gent13,walch14,hennebelle14}.

Figure~\ref{fig:hist}b shows how the temperature distribution changes if we weight by the mass of atomic hydrogen rather than by the total mass. At temperatures above a few hundred K, the gas is almost entirely atomic and so the temperature distribution is very similar to the one we recover when we weight by the total gas mass. At lower temperatures, however, there are clear differences between the two distributions. In both cases, there is a clear peak around $T \sim 100$~K, corresponding to the equilibrium temperature in the CNM, but there is much less gas at $T \ll 100$~K in the H-weighted distribution than in the distribution weighted by total gas mass. Figure~\ref{fig:hist}c, 
which shows an H$_{2}$ mass-weighted histogram of the temperature, demonstrates that this difference in behaviour is due to a change in the chemical composition of the gas. 
Very little of the H$_{2}$ in the simulation is located in gas with $T > 100$~K, but at lower temperatures, a significant fraction of the gas is molecular. 

Interestingly, Figure~\ref{fig:hist}c shows that the temperature distribution of the H$_{2}$ is relatively flat in the temperature range between $10 < T < 100$~K, with comparable amounts
of H$_{2}$ being found close to 10~K and close to 100~K.\footnote{It is possible that this flat distribution results from two broad, overlapping peaks of roughly equal height, similar to the peaks of unequal height we find in our Strong field simulation (Section~\ref{sec:change}),  but confirming this would require us to run additional simulations. In any case, whether the distribution is truly flat or broad and bimodal does not significantly affect our analysis or conclusions.} This contrasts strikingly with what we find if we instead weight the temperature distribution by the mass of CO, rather than the H$_{2}$ mass (Figure~\ref{fig:hist}d). The CO mass-weighted histogram shows a much narrower, strongly peaked temperature distribution centered around $T \sim 12$~K. Very little of the CO is located in regions with $T > 30$~K. 
This is consistent with observational determinations of the CO excitation temperature in typical Galactic GMCs, which recover values of around 10--30~K in regions not directly affected by stellar feedback \citep[e.g.][]{RomanDuval2010,poly12,nishi15}, although we caution that this only provides an accurate measure of the kinetic temperature if the CO emission is thermalized.
Comparison of Figure~\ref{fig:hist}c and Figure~\ref{fig:hist}d therefore clearly demonstrates that in addition to the expected CO-bright molecular gas, which we can loosely identify as the H$_{2}$ with $10 < T < 30$~K, there is also a CO-dark molecular component with $T > 30$~K that contains a significant fraction of the total molecular gas mass, consistent with the results presented in \citet{smith14}. 

\begin{figure}
\includegraphics[width=3.45in]{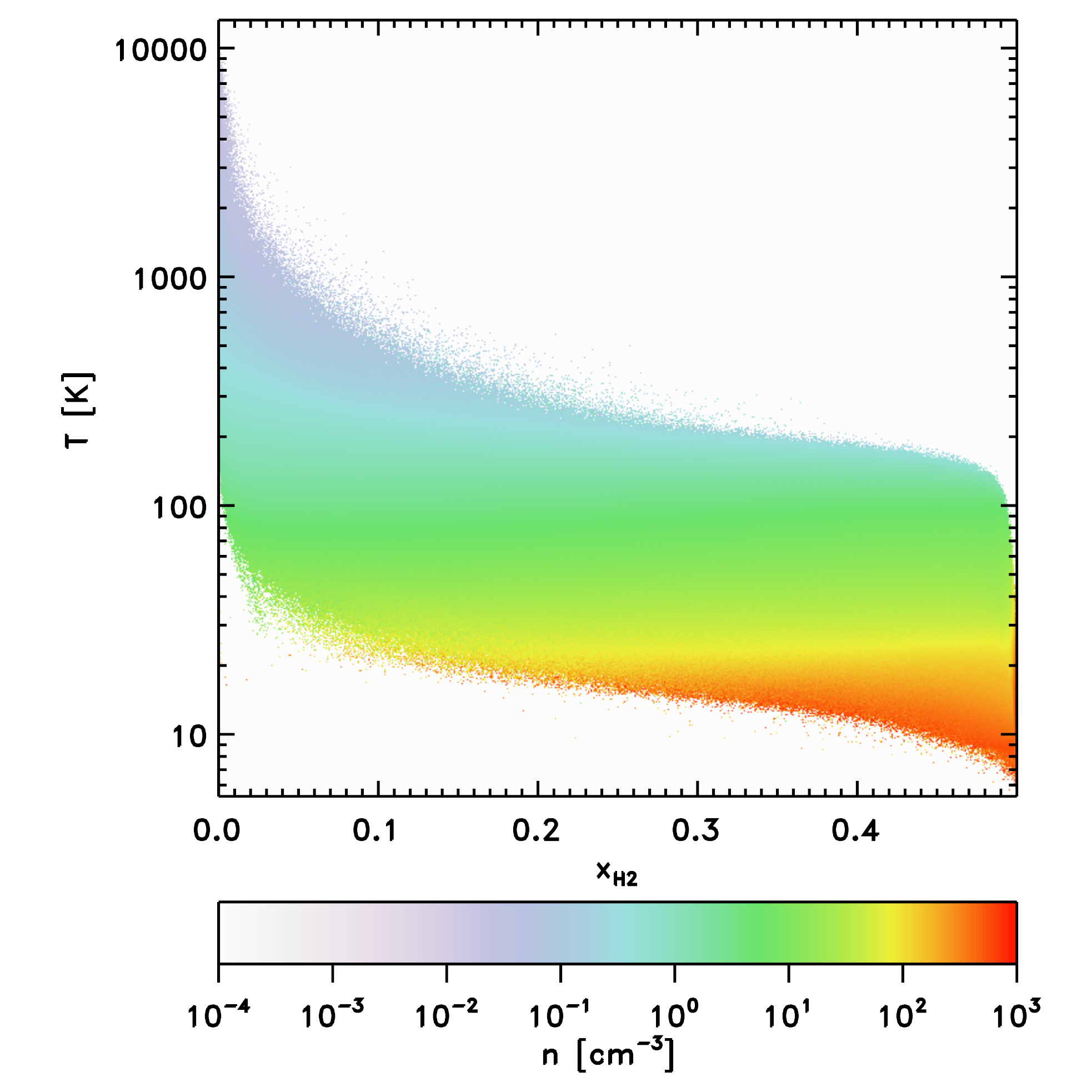}
\caption{Two-dimensional density-weighted probability distribution function showing the H$_{2}$ fractional abundance and gas 
temperature in the Milky Way simulation at $t = 261.1 \: {\rm Myr}$. \label{fig:2DPDF}}
\end{figure}

To help us to better characterize the nature of the gas in which we find H$_{2}$ but little or no CO, we have constructed a two-dimensional probability distribution function (PDF) for the H$_{2}$ fractional abundance and the gas temperature, weighted by the density of the gas. This is shown in Figure~\ref{fig:2DPDF}. 
We see from this Figure that the cold H$_{2}$ (i.e.\ the gas with $T < 30$~K) is almost all located in regions dominated by molecular hydrogen, and that this gas generally has a density 
$n > 100 \: {\rm cm^{-3}}$. On the other hand, the warmer H$_{2}$ (i.e.\ the component with $30 < T  < 100$~K) is found in regions spanning a wide range of different H$_{2}$ abundances, ranging from values as small as $x_{\rm H_{2}} \sim 0.01$ to as large as $x_{\rm H_{2}} \sim 0.5$.  Figure~\ref{fig:2DPDF} also shows us that the density of this gas is systematically lower than that of the colder, CO-bright gas, typically lying in the range $10 < n < 100 \: {\rm cm^{-3}}$. This range of densities corresponds more closely to those expected in the atomic CNM than those found in molecular clouds. This plot therefore helps to emphasize the fact that the physical conditions in the CO-dark molecular gas differ significantly from those in the CO-rich regions: the CO-dark gas is systematically warmer and less dense than the CO-rich gas, and so observational probes of the temperature and density of the latter component do not give a good description of the behaviour of the CO-dark gas.

\subsection{Comparison with previous numerical models}
The range of temperatures that we find in the CO-dark gas in our Milky Way simulation is in good agreement
with the results of previous theoretical models. For example, \citet{wolf10} study the composition and
properties of the CO-dark envelopes surrounding GMCs using a sophisticated 1D photodissociation region
(PDR) code. They find that the bulk of the CO-dark gas in their models has a temperature in the range
from 40--80~K. This is slightly narrower than the range we find in our simulation, but this may simply be
because their assumed densities are somewhat higher than the range of values that we find in our models.
\citet{wolf10} do not consider low extinction clouds that are not associated with bright CO emission, and so
make no predictions for this case.

Another recent theoretical study of the chemical and thermal distribution of the gas within a
representative CO-bright molecular cloud was carried out by \citet{offner13} using the three-dimensional
PDR code {\sc 3D-PDR} \citep{bisbas12}. They recover temperatures in the CO-dark component in the range
30--70~K. Again, they do not consider the case of a cloud without associated CO emission, and hence do
not probe the full range of conditions in which we find CO-dark molecular gas. 

In addition to these recent studies, there is also an extensive literature on the chemical and thermal
modelling of diffuse molecular and translucent clouds \citep[see e.g.\ the comprehensive review by][]{sm06}.
These clouds, observed in absorption, have large H$_{2}$ fractions but negligible CO, and hence represent
part of the CO-dark molecular gas component in the ISM. Models of these clouds \citep[e.g.][]{shaw06} again
recover characteristic temperatures in good agreement with those reported here.

Finally, dynamical models of the formation and evolution of individual molecular clouds that track the chemical and thermal 
evolution  of the gas \citep[e.g.][]{clark12b,gc12b,glover15} can also be used to study the properties of the CO-dark 
molecular phase, yielding results that once again agree well with those from our Milky Way simulation.

An important point to note, however, is that although all of these models (including our own) show that the
temperature of the CO-dark gas is sensitive to its density and extinction, most of the models are unable to
predict the distributions of densities and extinctions within the gas, and hence are unable to predict how 
the gas is distributed within the 30--100~K temperature range that all of the models largely agree on.
For example, in the PDR models, the density structure of the cloud is adopted as an input, rather than 
being an outcome of the modelling. The dynamical models discussed above do not suffer from this problem,
as the temperature and density in these models are evolved together in a self-consistent fashion. However,
in models that focus on individual clouds, there is always the concern that the results will be sensitive to the
assumed initial conditions for the cloud. To avoid this problem, it is necessary to carry out dynamical simulations
that evolve the ISM on scales much larger than the sizes of individual clouds, such as the disk models presented
in this paper.

\section{Changing the surface density and radiation field strength}
\label{sec:change}
In the previous section, we examined the temperature distribution of the molecular gas 
in our Milky Way simulation at a representative time during the evolution of the disk.
In this section, we extend our analysis to consider the other three simulations originally
presented in \citet{smith14}: the Low Density, Strong Field, and Low \& Weak simulations.
Doing so allows us to investigate how the temperature distribution of the CO-dark 
molecular gas depends on the mean surface density of the gas disk and the strength of
the interstellar radiation field. 

\begin{figure*}
\begin{overpic}[width=3.2in]{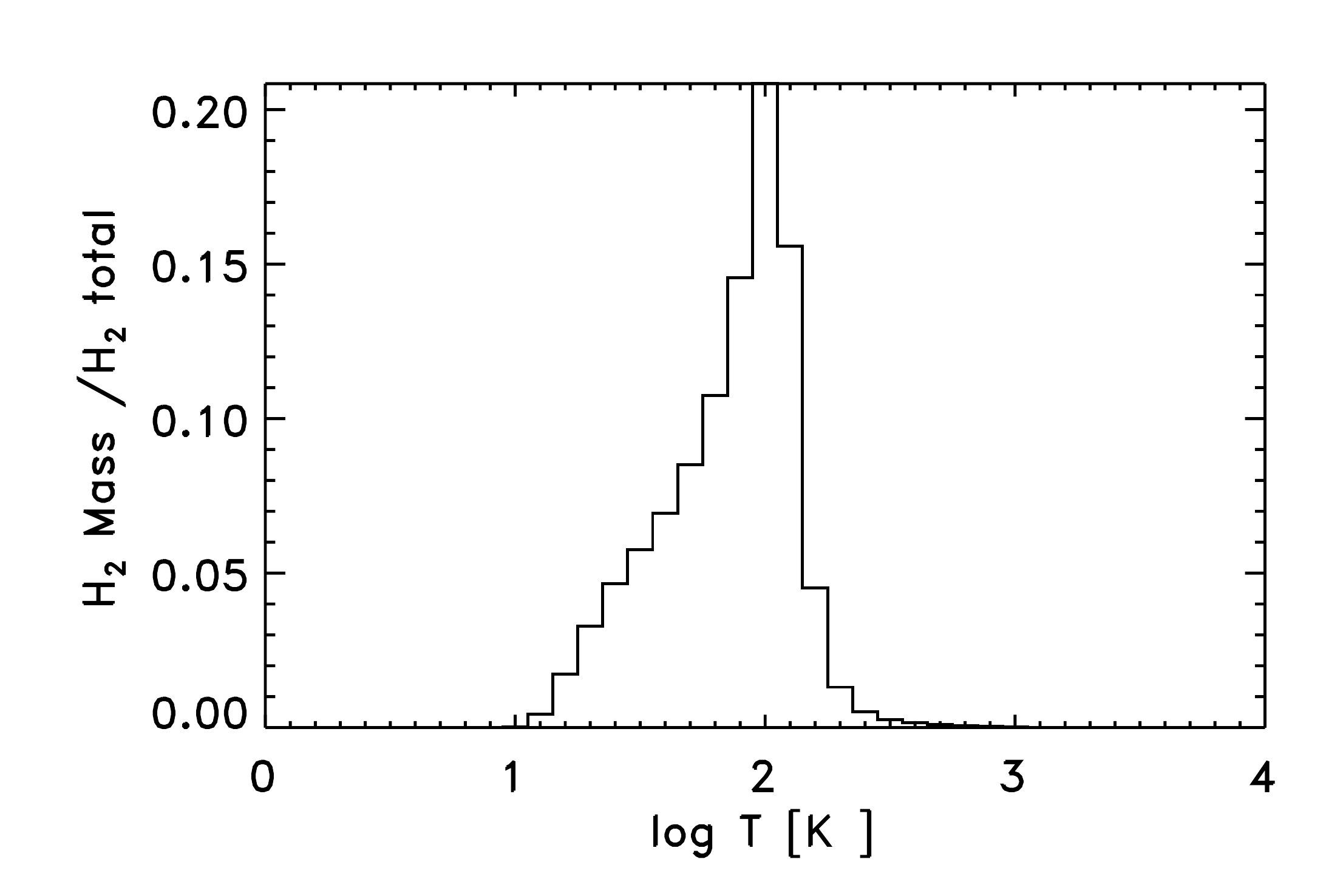}
\put(85,52){(a)}
\end{overpic}
\begin{overpic}[width=3.2in]{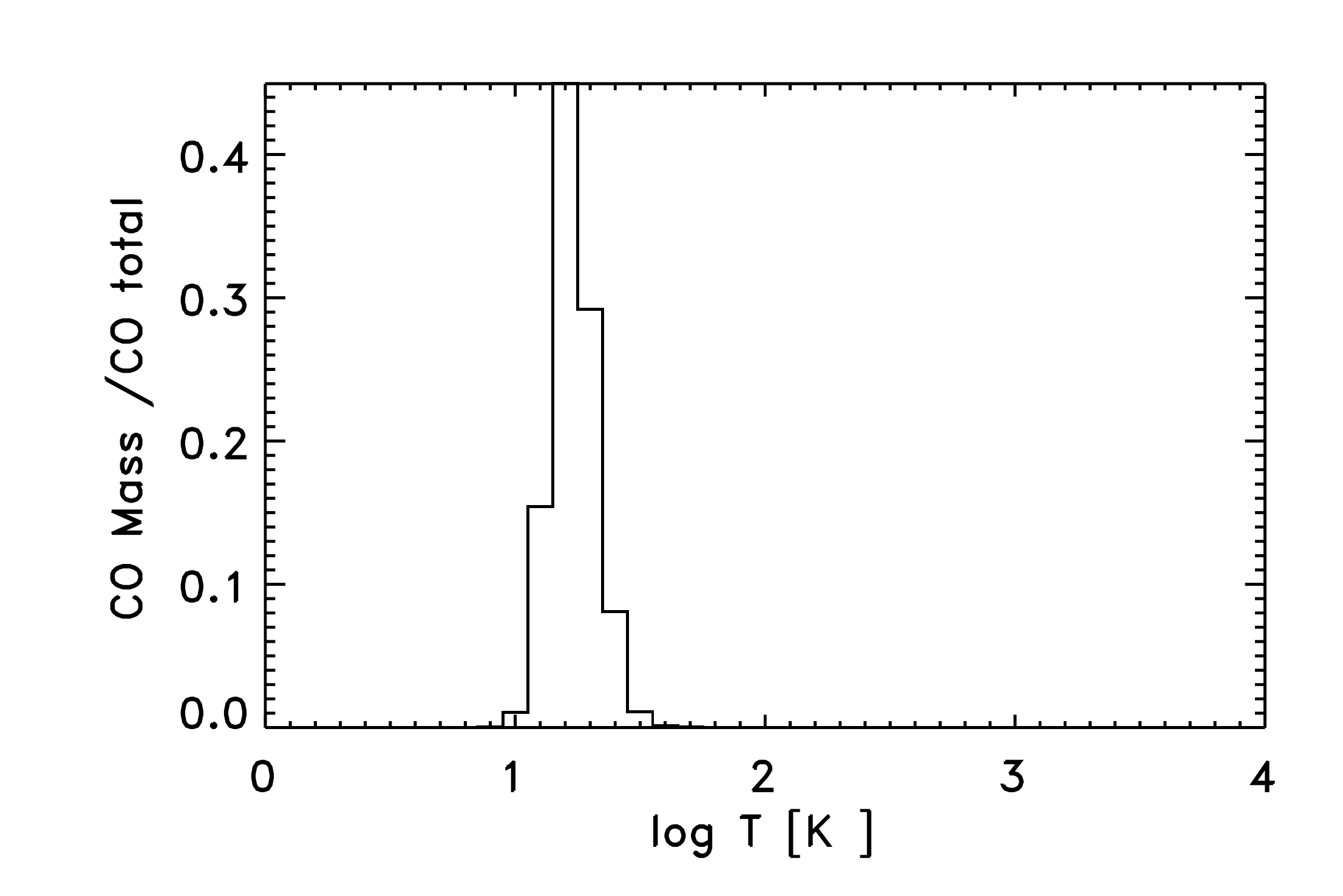}
\put(85,52){(b)}
\end{overpic}
\caption{(a) Histogram of the gas temperature in our Low Density simulation at time $t = 261.1 \: {\rm Myr}$, weighted by the H$_{2}$ mass.
(b) As (a), but weighted by the mass of CO.
\label{fig:hist_LD}}
\end{figure*}

In the Low Density simulation, the H$_{2}$ mass-weighted temperature distribution is qualitatively
different from the corresponding distribution in the Milky Way simulation. Instead of a relatively
flat distribution of temperatures between 10~K and 100~K, we find that the distribution is sharply
peaked around 100~K and falls off rapidly at both lower and higher temperatures. The high temperature
fall off occurs for the same reason that it does in the Milky Way simulation: most of the gas with $T > 100$~K 
has a density $n < 10 \: {\rm cm^{-3}}$, and the equilibrium H$_{2}$ fraction in this low density part of the
cold ISM is small. On the other hand, the fall off in the temperature distribution of the H$_{2}$ at low 
temperatures reflects the fact that only a small fraction of the gas in the low density disk is found in
clouds that have surface densities high enough to shield themselves against the effects of photoelectric
heating by the ISRF. This is consistent with our finding in \citet{smith14} that the dark gas fraction in this
simulation is very high, $\sim 90$\%, since the extinction required to shield the gas effectively against
CO photodissociation is similar to that required to shield it against photoelectric heating. 

If we focus on the small fraction of gas that does develop a significant CO content by examining
the CO mass-weighted temperature distribution (Figure~\ref{fig:hist_LD}b), then we see that in this
case there is much less difference between the Low Density run and the Milky Way run: in both
cases, the temperature distribution extends from around 10--30~K, with a peak in the middle of
this range. This demonstrates that the thermodynamical condition of the gas within a CO-bright
cloud is largely insensitive to the gas surface density of the galaxy averaged on much larger
scales, which is a straightforward consequence of the fact that local gas and dust shielding 
dominates in these clouds. 
 
\begin{figure*}
\begin{overpic}[width=3.2in]{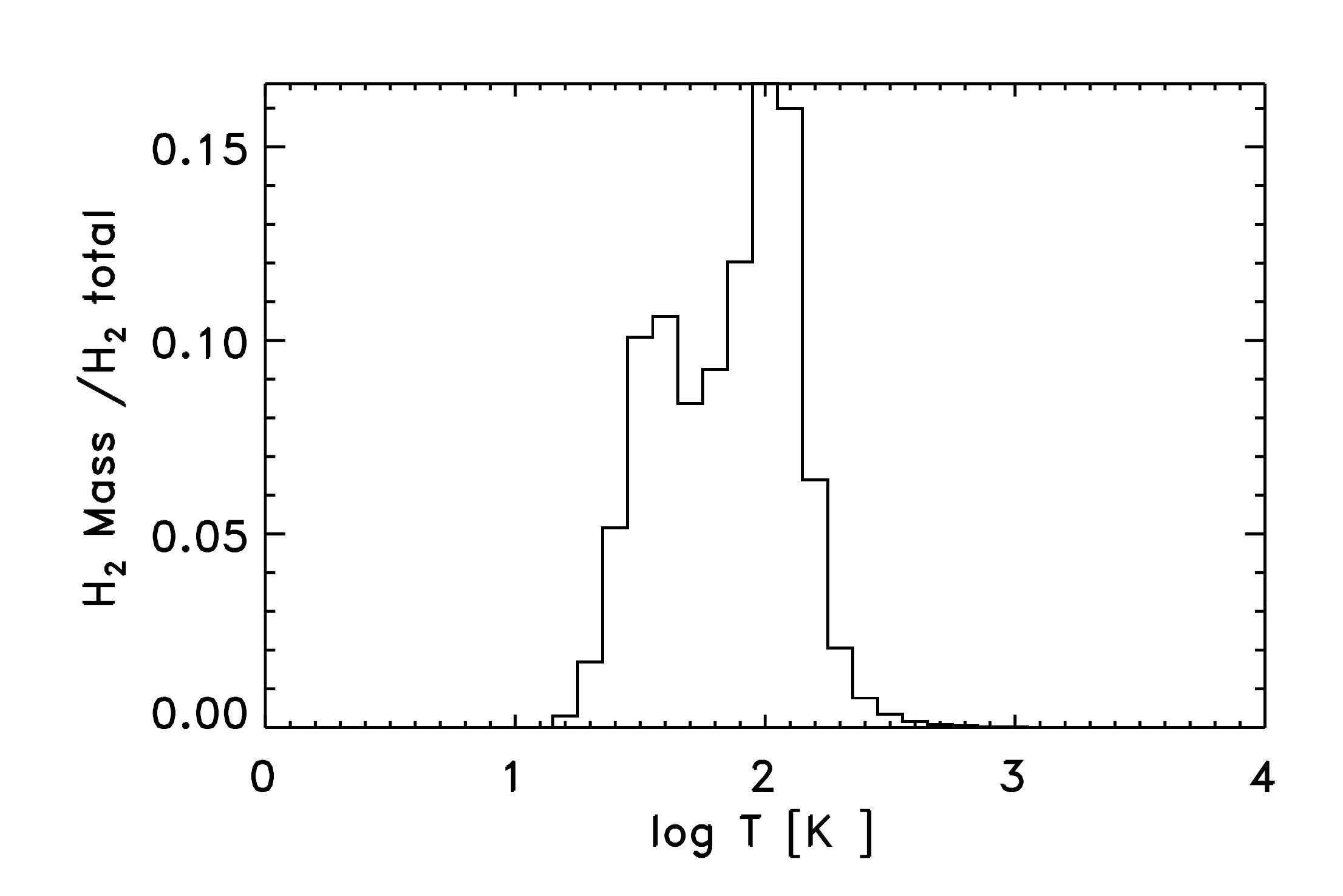}
\put(85,52){(a)}
\end{overpic}
\begin{overpic}[width=3.2in]{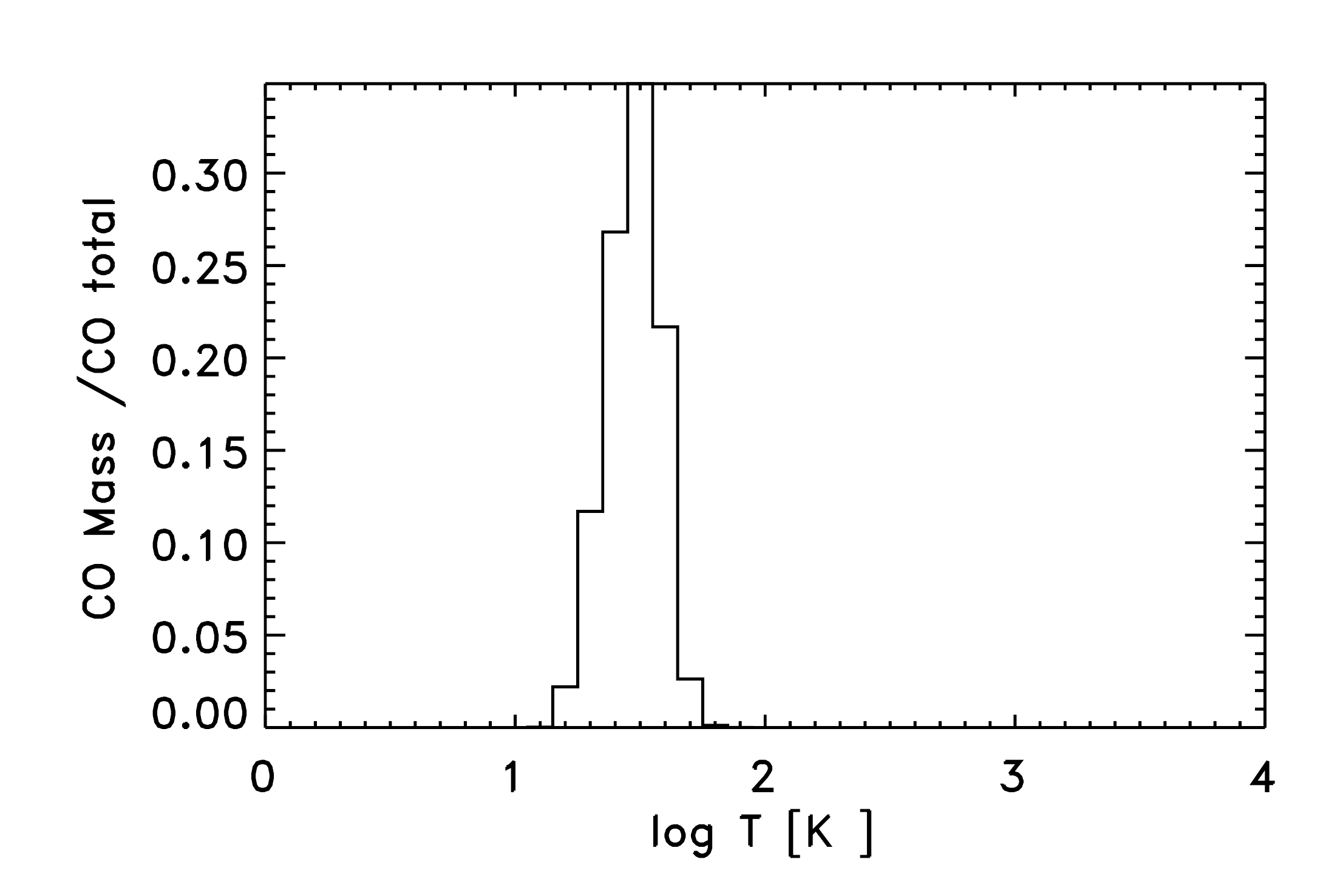}
\put(85,52){(b)}
\end{overpic}
\caption{(a) Histogram of the gas temperature in our Strong Field simulation at time $t = 261.1 \: {\rm Myr}$, weighted by the H$_{2}$ mass.
(b) As (a), but weighted by the mass of CO.
\label{fig:hist_SF}}
\end{figure*}

\begin{figure*}
\begin{overpic}[width=3.2in]{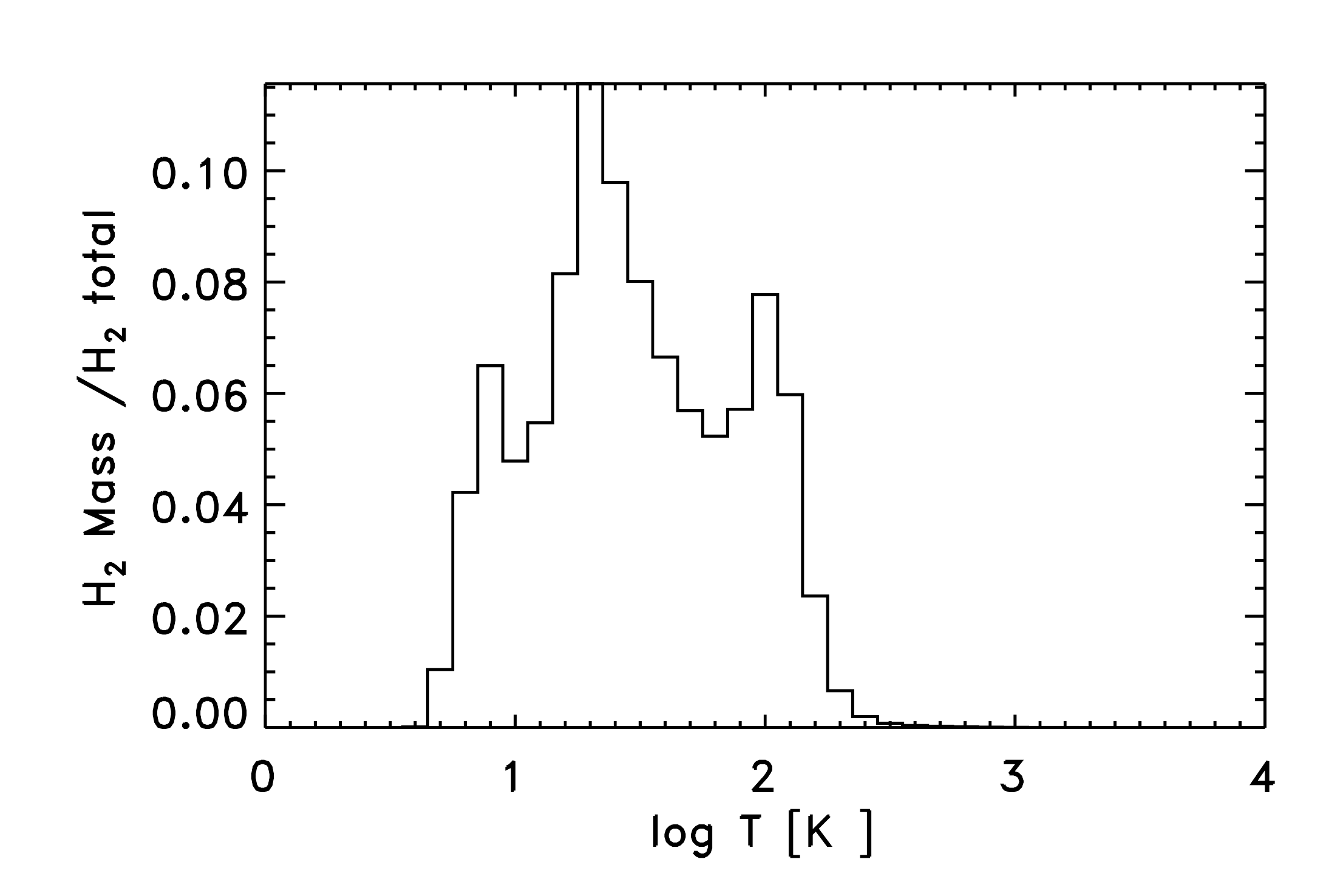}
\put(85,52){(a)}
\end{overpic}
\begin{overpic}[width=3.2in]{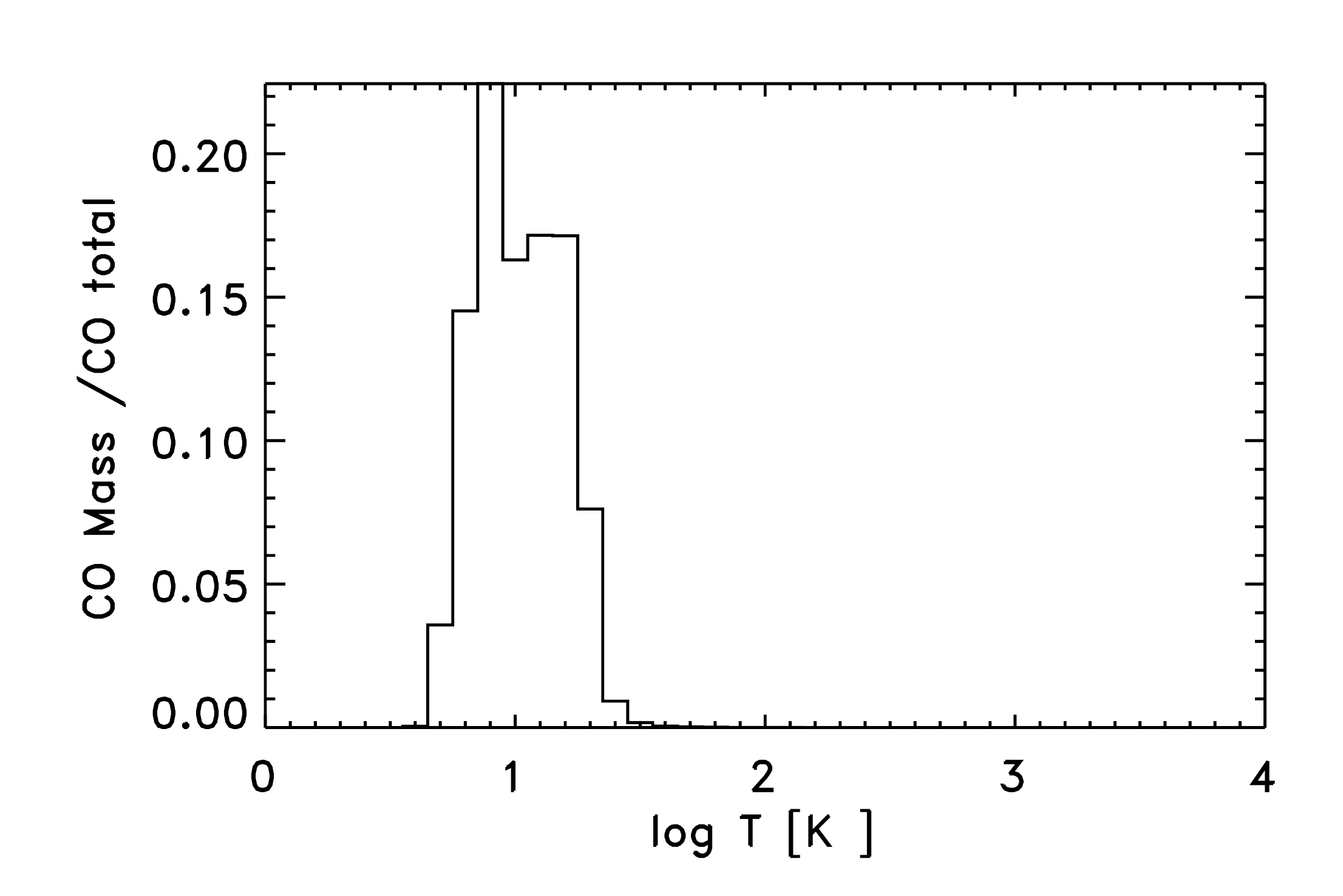}
\put(85,52){(b)}
\end{overpic}
\caption{(a) Histogram of the gas temperature in our Low \& Weak simulation at time $t = 261.1 \: {\rm Myr}$, weighted by the H$_{2}$ mass.
(b) As (a), but weighted by the mass of CO.
\label{fig:hist_LW}}
\end{figure*}

If we now look at the temperature distribution of the H$_{2}$ in the Strong Field simulation
(Figure~\ref{fig:hist_SF}a) we find that its behaviour lies in between that in the Milky Way simulation 
at that in the Low Density simulation. As in the Low Density simulation, the distribution is dominated 
by a peak close to $T \sim 100$~K, corresponding to CO-dark H$_{2}$ that is located in gas that is
efficiently heated by the stronger ISRF. However, the temperature distribution does not fall
off as rapidly as we move to lower temperatures as in the Low Density run, and there is a
clear second peak at $T \sim 30$~K, corresponding to the H$_{2}$ located in CO-bright
clouds. The CO-weighted temperature distribution is again narrow, with a single peak, but
is shifted to higher temperatures compared to its value in the Milky Way simulation, peaking
at around 30~K (see Figure~\ref{fig:hist_SF}b).

Finally, in the Low and Weak simulation, we  find that the H$_{2}$ has a broad 
and multiply-peaked distribution of temperatures, similar to the behaviour in the Milky Way
run. However, in this case, the main peak is located at low temperatures, $T \sim 25$~K,
owing to the low photoelectric heating rate in this simulation. The CO temperature distribution
is also shifted to lower values, peaking just below 10~K.

In summary, we see that the temperature distribution of the H$_{2}$ is highly sensitive to both the
strength of the ISRF and also the mean surface density of the gas in the disk. Simulations with high
values for the ISRF strength or low surface densities produce much more warm H$_{2}$ than cold
H$_{2}$. On the other hand, the run with the canonical Milky Way values for these parameters yields an
approximately flat H$_{2}$ temperature distribution, and the run with a low value for the ISRF strength
produces a distribution that is dominated by cold H$_{2}$. The temperature distribution of CO in the
simulations displays less variation, with all four runs producing a narrow distribution with a single
peak. However, the location of this peak does vary if we change the strength of the ISRF,
ranging from 8~K in the Low and Weak simulation to $33 \, $K in the Strong Field simulation. 
This is consistent with the observations of molecular clouds presented in \citet{RomanDuval2010} that show 
a systematic increase in the mean CO excitation temperature with decreasing Galactocentric radius 
(and hence increasing ISRF strength),  although we caution that this might also reflect a change in the mean
density of the clouds.
 
\section{Can we map CO-dark molecular gas using \cii and \oino?}
\label{sec:map}
\subsection{Generating maps of \cii and \oi emission}
As we have seen in the previous sections, the characteristic temperature of
CO-dark molecular gas is significantly higher than that of the CO-rich gas
making up most GMCs. This raises the prospect of detecting this CO-dark
molecular component using the fine structure lines of ionized carbon and
neutral oxygen. 

The C$^{+}$ ground state is split into two fine structure
levels, $^{2}P_{1/2}$ and $^{2}P_{3/2}$, separated by an energy 
$E_{10} / k = 92$~K, where $k$ is
Boltzmann's constant. For O, we have instead three fine structure levels
in the ground state, $^{3}P_{2}$, $^{3}P_{1}$ and $^{3}P_{0}$, 
with energy separations of $E_{10} / k = 227$~K between $^{3}P_{2}$ 
(the lowest energy level) and $^{3}P_{1}$ and $E_{20} / k = 327$~K between 
$^{3}P_{2}$ and $^{3}P_{0}$.
At typical GMC temperatures of 10--30~K, it is difficult
to excite any of the excited levels of C$^{+}$ or O. For example, the exponential factor in the
collisional excitation rate of C$^{+}$ has a value of $e^{-9.2} \simeq 10^{-4}$
at 10~K and $e^{-9.2/3} \simeq 0.05$ at 30~K. Emission from the \oi
lines is suppressed even more strongly at these temperatures. However,
at the higher temperatures found in the CO-dark molecular gas, the
situation is more promising: at 60~K, for instance, the \cii excitation rate
is suppressed by a factor of only $e^{-9.2/6} \sim 0.2$. 

For this reason, it is interesting to compute how bright the \oi and \cii
emission from the CO-dark molecular component is likely to be, in order
to allow us to assess how easy it will be to detect the CO-dark H$_{2}$
using these lines. The obvious way in which to do this would be to take
the results of our simulation and run them through a full radiative transfer
code, as we have done with many of our smaller-scale simulations in
previous papers \citep[e.g.][]{smith13,glover15}. Unfortunately, the sheer size of
the simulations studied in this paper renders this impractical: none of
the line radiative transfer codes to which we have access can handle 
problems with hundreds of millions of discrete resolution elements and
dynamical ranges that run from $\sim 0.3$~pc (the best spatial resolution
we achieve in the highest resolution parts of the disk) up to a few kpc.

Fortunately, it is possible to get a reasonably good idea of the likely
strength of the \cii and \oi emission without needing to solve the
full radiative transfer problem, simply by assuming that the fine
structure lines are optically thin. In this limit, we can calculate the
total integrated brightness of the \cii 158$\,\mu$m line or the 
\oi 63$\,\mu$m and 145$\,\mu$m lines along a given line of sight
simply by summing up the total number of photons emitted in these
lines in the direction of interest, without needing to worry about whether
any of these photons are later absorbed. The conditions in which this
assumption is likely to be valid are discussed in Appendix~\ref{app:opthick}.

Having made the assumption that the \cii and \oi lines are optically thin,
we can compute the amount of energy radiated in each line per unit volume
by assuming that the fine structure levels are in statistical equilibrium and
solving the appropriate set of algebraic equations for the level populations.
For example, in the case of \ciino, if we denote the fractional population of 
the ground-state as $x_{0}$ and the fractional population of the excited
state as $x_{1}$, then 
\begin{equation}
(A_{10} + B_{10} J_{10} + C_{10}) x_{0} = (B_{01} J_{10} + C_{01}) x_{1},
\end{equation}
where $A_{10}$, $B_{10}$ and $B_{01}$ are the Einstein coefficients for
spontaneous emission, stimulated emission and absorption for
the $1 \rightarrow 0$ transition, respectively, $J_{10}$ is the mean specific 
intensity of the incident radiation field at the wavelength of the line,
and $C_{01}$ and $C_{10}$ are the collisional excitation and de-excitation
rates, respectively. Since we also know that $x_{0} + x_{1} = 1$, it is simple
to derive the following expression for $x_{1}$:
\begin{equation}
x_{1} = \frac{A_{10} + B_{10} J_{10} + C_{10}}{A_{10} + (B_{10} + B_{01}) J_{10} + C_{10} + C_{01}}.
\end{equation}
To solve this, we need to specify the various excitation and de-excitation 
rates, as well as the value of $J_{10}$. For $A_{10}$ we adopt the
value $A_{10} = 2.3 \times 10^{-6} \: {\rm s^{-1}}$ specified in the {\sc lamda} 
database \citep{sch05}; $B_{01}$ and $B_{10}$ then follow from this via the standard 
relationships.  Our value for $J_{10}$ is taken from \citet{mmp83}. In our Milky Way 
simulation, we interpolate the value at $158 \: \mu$m from their tabulated values for
the field strength in the wavelength range 8--1000~$\mu$m for the solar neighbourhood
(see their Table B1). For our other simulations, we assume that the field strength at this frequency scales
linearly with the UV field strength -- i.e.\ we adopt a value ten times larger in the Strong
Field simulation and ten times weaker in the Low and Weak simulation. Since we expect
the ISRF at  $158 \: \mu$m to be dominated by dust emission rather than by the CMB,
this simple rescaling procedure is appropriate, at least at the level of accuracy of the
calculations reported on here.
For $C_{10}$, we
account for collisions between C$^{+}$ and electrons, hydrogen atoms and
H$_{2}$ molecules, using rate coefficients taken from \citet{wb02}, 
\citet{hm89} and \citet{wg14}, respectively, and values for the number densities
of ${\rm e^{-}}$, H and H$_{2}$ taken from the simulations.
Finally, $C_{01}$ follows from $C_{10}$ if we apply the principle of detailed balance. 

Given $x_{1}$, we can then compute the frequency-integrated emissivity for
the \cii line,
\begin{equation}
j_{\rm CII} = \frac{1}{4\pi} (A_{10} + B_{10} J_{10}) E_{10} x_{1} n_{\rm C^{+}},
\end{equation}
where $n_{\rm C^{+}}$ is the number density of C$^{+}$ ions. However, as the
{\sc arepo} cells do not have a fixed volume, in practice it is easier to work in
terms of the emissivity per unit mass
\begin{equation}
\frac{j_{\rm CII}}{\rho} = \frac{(A_{10} + B_{10} J_{10}) E_{10} x_{1} n_{\rm C^{+}}}{4\pi \rho}.
\end{equation}

The procedure in the case of the two \oi lines is very similar, except that
in this case we have to solve three coupled linear equations for the
fractional level populations of the ground state and the two excited states.
As with \ciino, we take the required atomic data from the {\sc lamda} 
database. When computing the collisional excitation and de-excitation
rates, we consider the effects of collisions with electrons, protons, atomic
hydrogen and H$_{2}$, using data from \citet{bbt98}, \citet{p90}, \citet{akd07} 
and \citet{j92}, respectively. We also account for the effects of charge
transfer with H$^{+}$
\begin{equation}
{\rm O + H^{+} \rightarrow O^{+} + H}
\end{equation}
followed by the inverse process
\begin{equation}
{\rm O^{+} + H \rightarrow O+ H^{+}}.
\end{equation}
This pair of reactions can leave the oxygen atom in a different fine structure 
state from the one that it started in, with the result that these reactions can in
some cases significantly affect the oxygen fine structure level populations
\citep{p90}. To model the effects of these reactions we make use of the fine-structure 
resolved rate coefficients for both processes computed by \citet{stan99}.

Finally, given the emissivity per unit mass for the \cii line for each
{\sc arepo} cell, we can compute the total \cii emissivity for the cell
simply by multiplying by the mass contained in the cell. The resulting
distribution of emissivities can be analyzed directly, and we can also
make maps of the emission by integrating along selected lines of sight.
The same procedure can also be used for the two \oi lines.

It is important to note that because our simulations do not include star formation
or stellar feedback, we are unable to model the \cii or \oi emission directly associated
with regions of massive star formation. In reality, some of the \cii observed in real
galaxies is produced within H$\,${\sc ii} regions, and a further substantial fraction
is produced within the bright PDRs surrounding massive O and B-type stars. These
PDRs also generate much of the observed \oi emission. Our synthetic images do not
account for these sources and this should be borne in mind when interpreting our
results.

\subsection{Analysis}
\subsubsection{Results from the Milky Way simulation}
\begin{figure}
\begin{overpic}[width=0.5\textwidth]{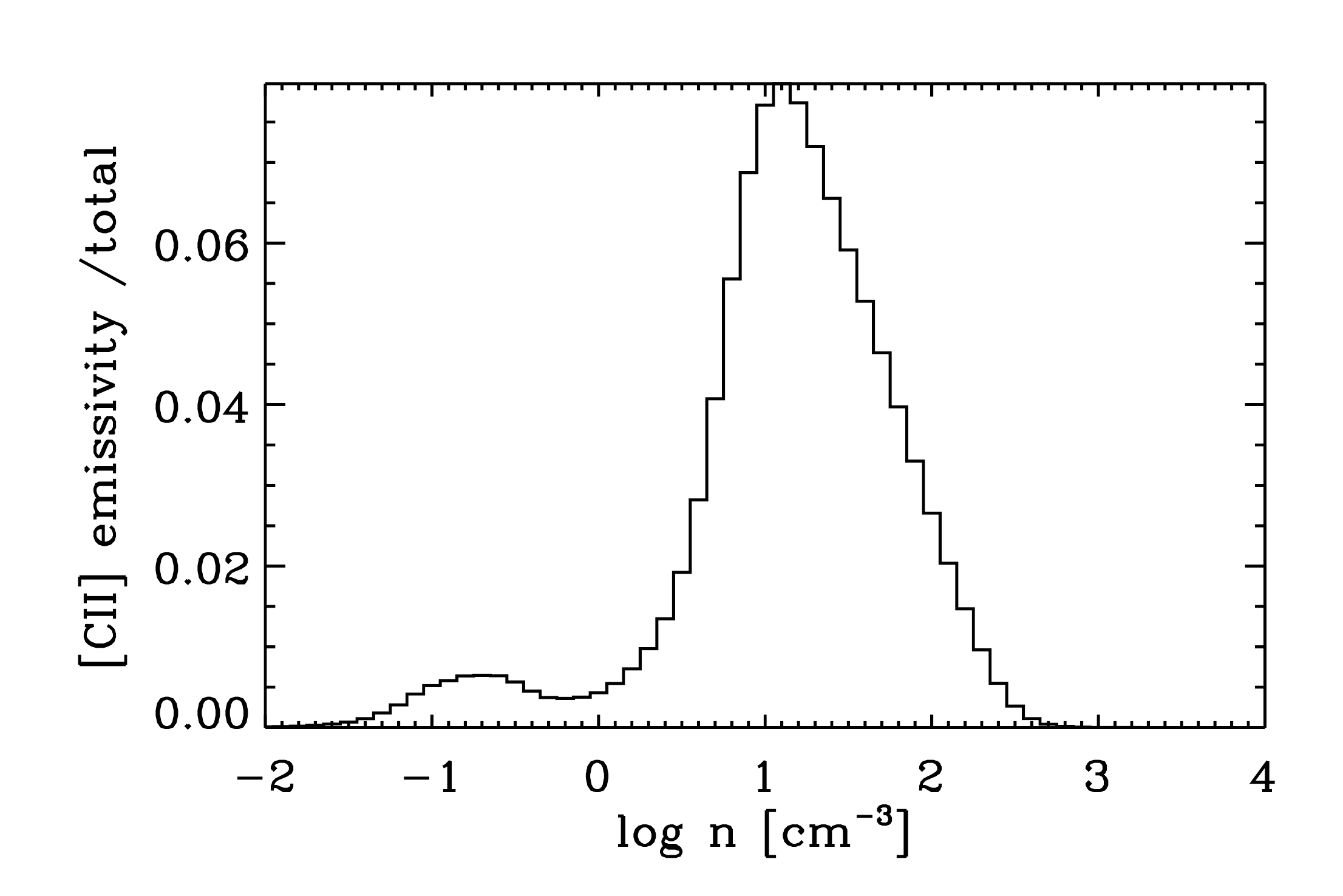}
\put(85,52){(a)}
\end{overpic}
\begin{overpic}[width=0.5\textwidth]{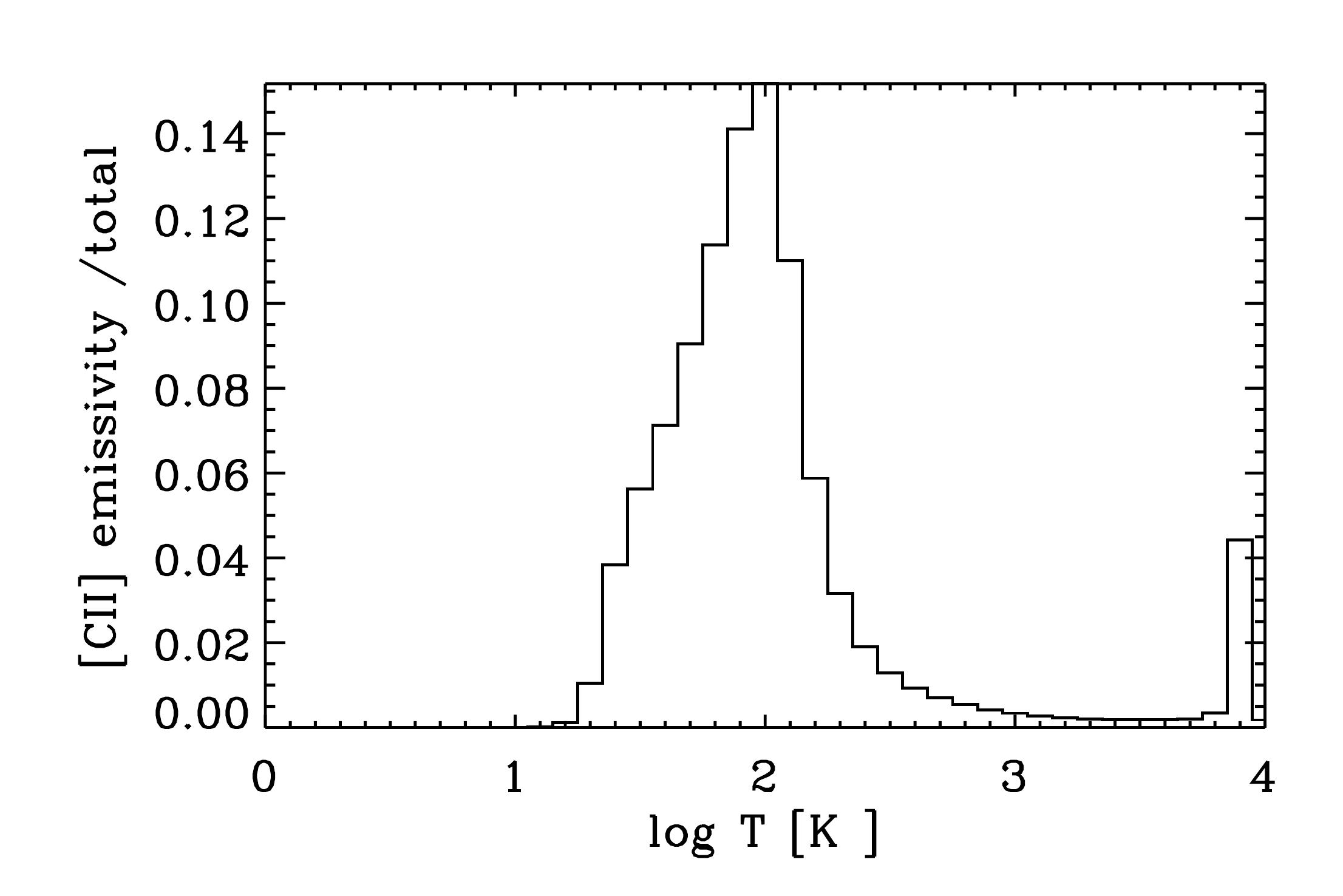}
\put(85,52){(b)}
\end{overpic}
\caption{(a) Distribution of the estimated \cii emissivity as a function of the gas number density in our
Milky Way simulation at time $t = 261.1 \: {\rm Myr}$. 
(b) As (a), but showing the distribution with respect to the gas temperature.
\label{fig:cphist}}
\end{figure}

In Figure~\ref{fig:cphist}, we show how the estimated \cii emissivity is
distributed as a function of density and temperature in the Milky Way
simulation at time $t = 261.1 \: {\rm Myr}$.
The distribution with respect to number density is bimodal,
with a small peak centered on $n \sim 0.2 \: {\rm cm^{-3}}$, corresponding
to emission from the WNM, and a much larger peak at $n \sim 20 \:
{\rm cm^{-3}}$, corresponding to emission from the cold atomic and
CO-dark molecular gas. The distribution with respect to temperature
tells us a similar story, with a sharp peak just below $10^{4} \: {\rm K}$
corresponding to the emission from the WNM and a broader peak around
100~K corresponding to emission from the cold atomic and molecular gas.

We saw in Section~\ref{sec:tempd} that the temperature distribution of the H$_{2}$
in the Milky Way simulation is relatively flat between $T \sim 10$~K and
$T \sim 100$~K. This is not the case with the \cii emissivity, which
decreases sharply with decreasing temperature, owing to the influence
of the $e^{-92/T}$ term in the collisional excitation rate. It is therefore
clear that the \cii emission cannot be tracing all of the H$_{2}$ equally well.
Indeed, if we compare the temperature distribution of the \cii emission
with the CO mass-weighted temperature distribution shown in Figure~\ref{fig:hist}d,
we see that there is little overlap: \cii is a very poor tracer of the cold, 
dense CO-bright clouds, but performs much better as a tracer of warmer,
more diffuse gas. 

\begin{figure}
\includegraphics[width=0.5\textwidth]{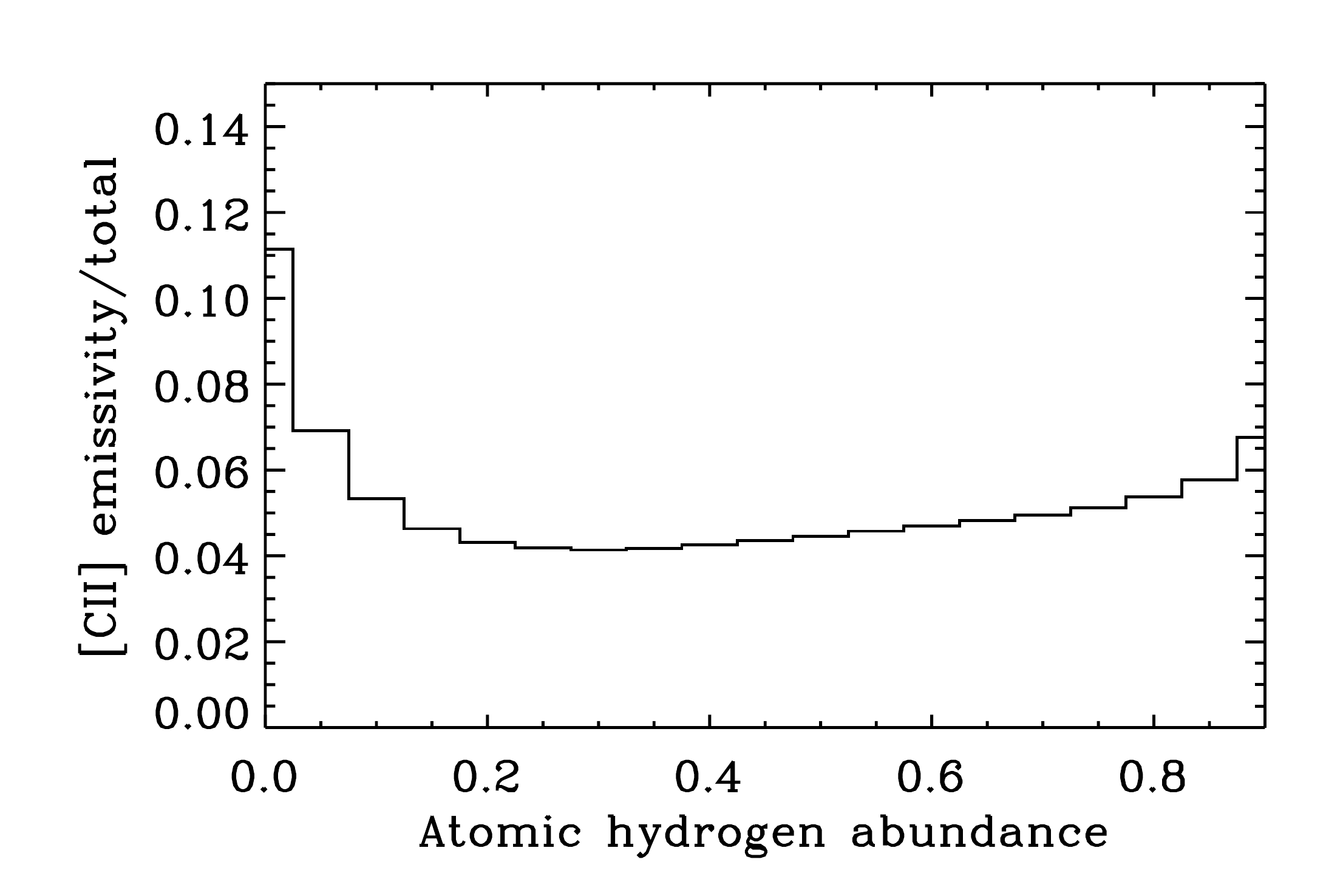}
\caption{As Figure~\ref{fig:cphist}a, but showing the distribution of
\cii emissivity as a function of the fractional abundance of atomic hydrogen.
\label{fig:cphist2}}
\end{figure}

There remains the question of how much of the total \cii emission 
in our simulations is produced by CO-dark molecular gas and how much
comes instead from atomic gas. To address this, we have constructed
a histogram of the \cii emissivity as a function of the fractional abundance
of atomic hydrogen, shown in Figure~\ref{fig:cphist2}. We see that there
is a clear peak close to $x_{\rm H} \sim 0$, corresponding to emission
from gas which is almost entirely molecular. However, there is also
substantial emission coming from regions with atomic hydrogen fractions
significantly greater than zero. If we define H$_{2}$-dominated regions
to be those with $x_{\rm H} < 0.5$ (i.e.\ with more than half of their total
hydrogen in the form of H$_{2}$), and H-dominated regions to be
those with $x_{\rm H} \geq 0.5$, then we find that around half of the total
\cii emission is produced in H$_{2}$-dominated gas, while the other half is
produced in H-dominated gas. For comparison, we note that \citet{vl14}
find that in our own Galaxy, the ratio of the total \cii emission produced
by H$_{2}$-dominated diffuse gas (i.e.\ by H$_{2}$ in regions which are
not PDRs) to that produced by H-dominated gas is roughly 57\% to 
43\%, in fairly good agreement with our results. However, 
at this point, it is worthwhile reminding the reader that our simulations
do not account for the effects of stellar feedback. Our estimates of
the \cii emission therefore do not account for the contributions coming
from H$\,${\sc ii} regions and the bright PDRs around massive O and
B stars, which together are responsible for roughly 75\% of the total
\cii emission of the Milky Way \citep{vl14}. Our
estimates of the relative fractions of the \cii emission produced by 
H$_{2}$-dominated and H-dominated gas therefore only apply to the
diffuse emission powered by the mean ISRF and not to the emission
coming from these brighter regions. 

\begin{figure}
\includegraphics[width=0.5\textwidth]{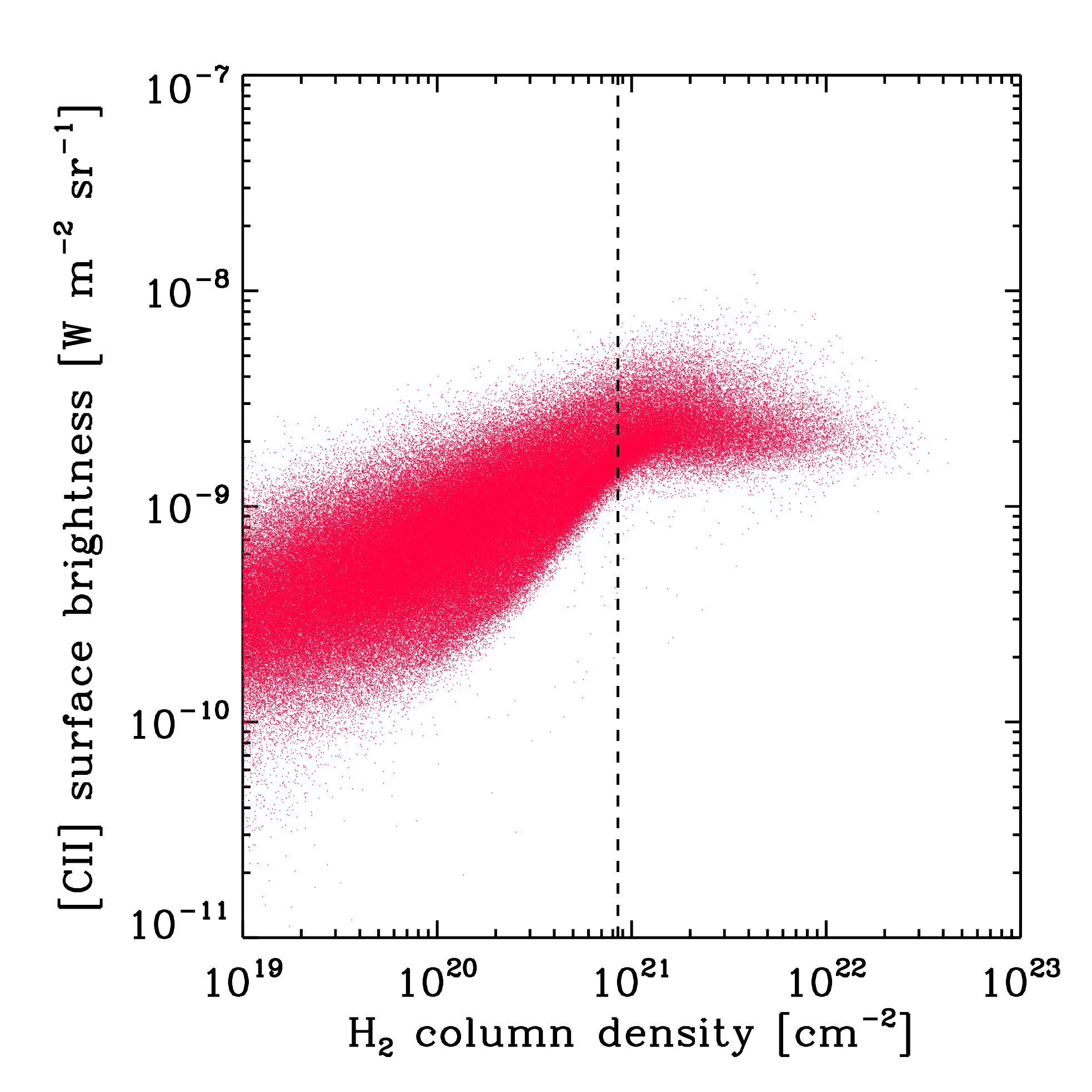}
\caption{Estimated \cii surface brightness, plotted as a function of
the H$_{2}$ column density, for the Milky Way simulation at
time $t = 261.1 \: {\rm Myr}$. The vertical dashed line indicates 
the H$_{2}$ column density at which we would expect the \cii
line to become optically thick, computed assuming a \cii 
1D velocity dispersion $\sigma_{\rm 1D} = 1 \: {\rm km \: s^{-1}}$ (see Appendix A). 
For larger linewidths, this column density scales as $N \propto \sigma_{\rm 1D}$. 
Note that as our
calculation of the \cii surface brightness does not account for the
line opacity, the flattening that we see in the relationship between
$I_{\rm CII}$ and $N_{\rm H_{2}}$ is not an optical depth effect.
Rather, it is a chemical effect: at higher $N_{\rm H_{2}}$, most of
the carbon is in the form of CO, and so we just see a constant
\cii surface brightness coming from the ``skin'' of the cloud.
\label{fig:LCII_NH2}}
\end{figure}

It is also important to remind the reader that our estimate here
assumes that \cii remains optically thin. In practice, many of the
lines of sight with the highest H$_{2}$ column densities lie in the
regime where \cii is likely to become optically thick (see 
Figure~\ref{fig:LCII_NH2}), and so it is possible that we are
overestimating the \cii emission from the highly molecular gas
to some degree. Note, however, that even if the \cii line is
optically thick, if the gas density is much lower than the critical
density for the transition, the \cii line can be ``effectively'' optically thin 
\citep{gold12}. In this case, the \cii photons produced will scatter but 
most will eventually escape, and so in this regime the total \cii luminosity 
is relatively insensitive to the \cii optical depth. \citet{gold12} show that
for H$_{2}$-dominated gas, $n_{\rm cr} = 6100 \: {\rm cm^{-3}}$ at
100~K and is even larger at lower temperatures. If we compare this
value with the H$_{2}$ number densities shown in Figure~\ref{fig:2DPDF}, 
we see that most of the molecular gas in our simulations is indeed in this
effectively optically thin regime.

Our results show that a significant fraction of the diffuse \cii emission
produced in the Milky Way simulation is produced by the CO-dark
molecular gas component. However, this only accounts for around
half of the total diffuse emission, with the remaining half coming from
regions dominated by atomic gas. 
Separating these two contributions -- essential if we are to use the
\cii emission as a tracer of the CO-dark gas -- is unlikely to be 
straightforward. In addition, the emissivity per unit mass changes
significantly over the range of densities and temperatures occupied
by the CO-dark gas. As a result, the relationship between the \cii 
surface brightness and the H$_{2}$ column density is not linear,
as illustrated in Figure~\ref{fig:LCII_NH2}. For lines-of-sight
with $N_{\rm H_{2}} < 10^{21} \: {\rm cm^{-2}}$, corresponding to
visual extinctions below $A_{\rm V} \sim 1$, the \cii surface brightness
scales with $N_{\rm H_{2}}$ approximately as $I_{\rm CII} \propto
N_{\rm H_{2}}^{1/2}$, although the scatter around this relationship
is considerable. At higher H$_{2}$ column densities, however, this
relationship breaks down. In this regime, we are probing dense,
CO-bright clouds, and so increasing the H$_{2}$ column density
merely increases the CO column density, but does not significantly
affect the amount of C$^{+}$ along the line-of-sight. We therefore
find that at high $N_{\rm H_{2}}$, the \cii surface brightness becomes
approximately constant. Essentially, what we are seeing is the
emission from the C$^{+}$-rich ``skin'' of the dense molecular cloud.

\begin{figure}
\includegraphics[width=0.5\textwidth]{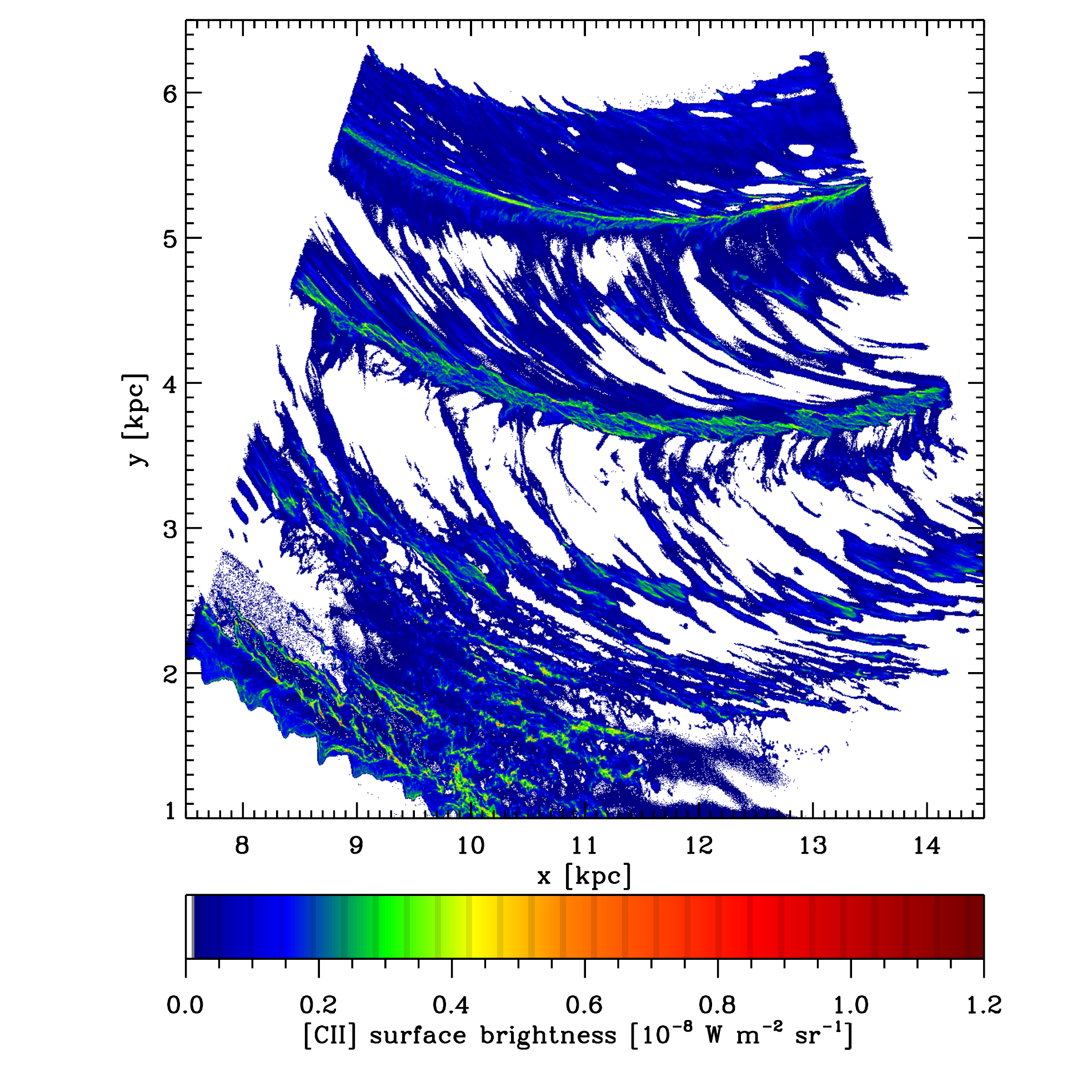}
\caption{Map of the estimated \cii surface brightness in our Milky Way
simulation at time $t = 261.1 \: {\rm Myr}$. Regions shown in white
have surface brightnesses $I_{\rm CII} \leq 10^{-10} \: {\rm W
\, m^{-2} \, sr^{-1}}$.
\label{fig:CII_SB}}
\end{figure}

A further problem with using \cii emission to trace CO-dark
molecular gas in the ISM conditions modelled in our Milky Way
simulation concerns the detectability of the resulting emission.
In Figure~\ref{fig:CII_SB}, we show a map of the \cii surface brightness
as seen by an observer looking straight down onto the disk.
We see that even in the brightest regions, the \cii surface
brightness does not exceed $1.2 \times 10^{-8} \:
{\rm W \: m^{-2} \: sr^{-1}}$, and that much of the CO-dark
H$_{2}$ is traced by emission with a \cii surface brightness of
only a few times $10^{-9} \: {\rm W \: m^{-2} \: sr^{-1}}$.
For comparison, maps of \cii emission from nearby spiral
galaxies made using the PACS instrument on board the
{\it Herschel} satellite typically have 3$\sigma$ detection
thresholds of around a few times $10^{-9} \: {\rm W \: m^{-2} \: sr^{-1}}$
to $10^{-8} \: {\rm W \: m^{-2} \: sr^{-1}}$ \citep{croxall12,parkin13}.
With the FIFI-LS instrument on board the SOFIA observatory,
it may be possible to go slightly deeper in large-scale maps,
to around $10^{-9} \: {\rm W \: m^{-2} \: sr^{-1}}$ (J.~Pineda,
private communication). Therefore, we predict that it should
be possible to trace the CO-dark H$_{2}$ in galaxies similar
to the Milky Way using large-scale \cii maps, but that doing 
so at high accuracy will be challenging.

\begin{figure}
\begin{overpic}[width=0.5\textwidth]{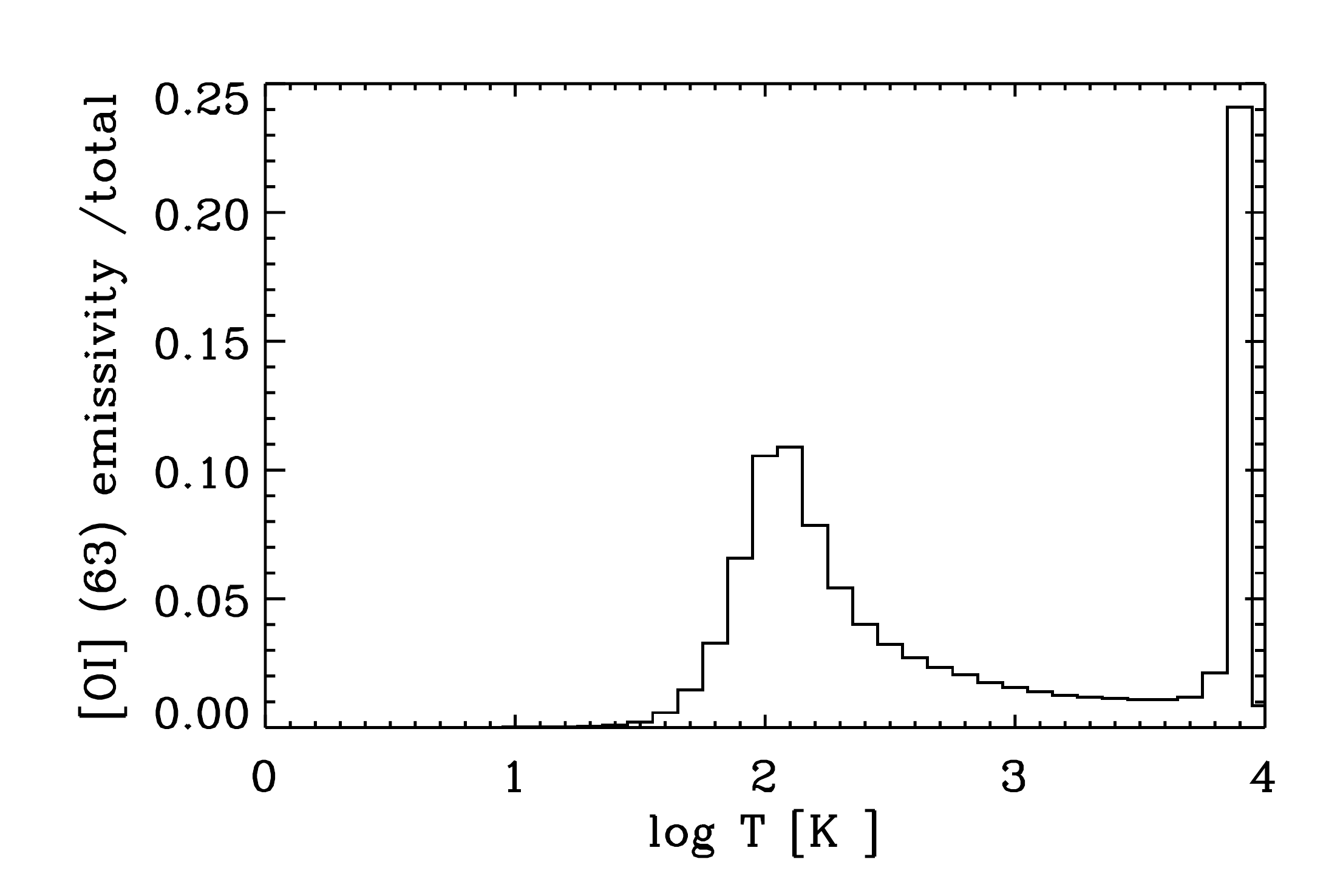}
\put(25,52){(a)}
\end{overpic}
\begin{overpic}[width=0.5\textwidth]{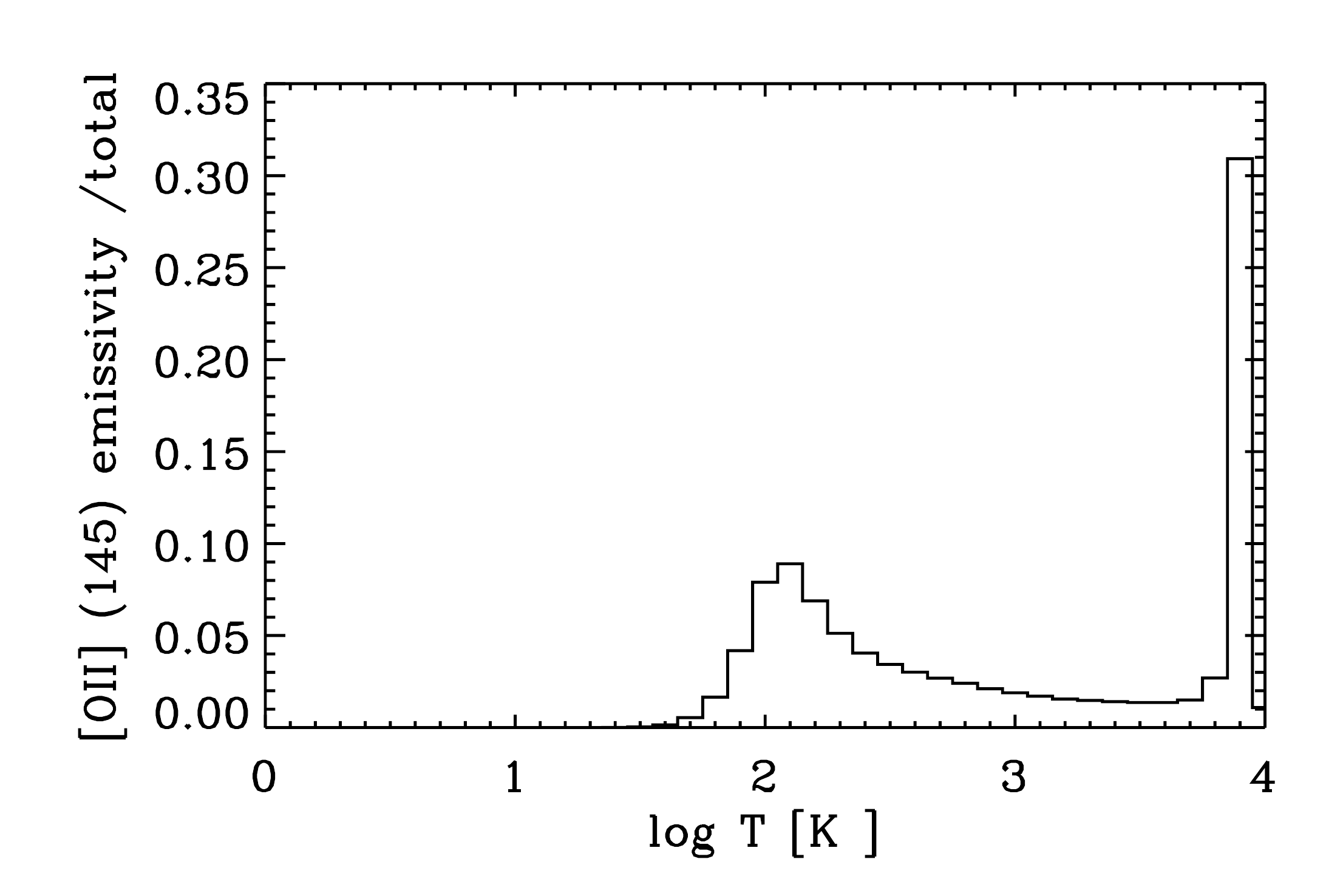}
\put(25,52){(b)}
\end{overpic}
\caption{(a) Distribution of the estimated \oi 63$\,\mu$m emissivity as a function of the 
gas temperature in our Milky Way simulation at time $t = 261.1 \: {\rm Myr}$. 
(b) As (a), but for the 145$\,\mu$m line of \oino.
\label{fig:ohist}}
\end{figure}

\begin{figure}
\includegraphics[width=0.5\textwidth]{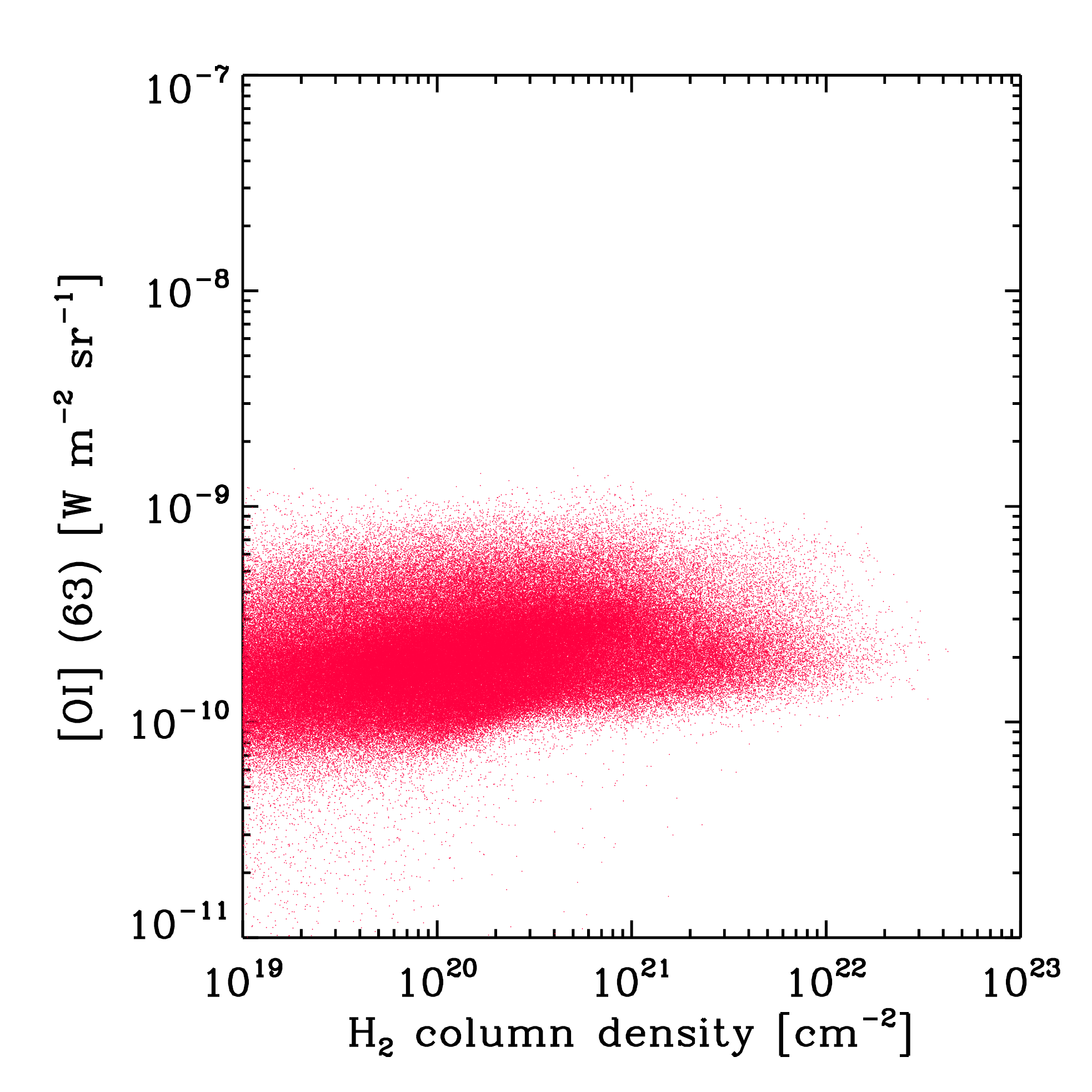}
\caption{As Figure~\ref{fig:LCII_NH2}, but for the \oi
63$\,\mu$m line. \label{fig:LOI_NH2}}
\end{figure}

If \cii is a problematic tracer of CO-dark molecular gas in
Milky Way-like conditions, then what about \oino? In this
case, the greater energy required to excite the fine
structure lines makes \oi a much worse tracer of the 
CO-dark molecular component than \ciino. This is illustrated
clearly in Figure~\ref{fig:ohist}, where we show the density distribution
of the estimated \oi emissivity in the 63$\,\mu$m and
145$\,\mu$m lines. In both cases, the majority of the
emission comes from gas with $T > 100$~K, which we
already know is mostly atomic. Only 31.4\% of the total
emission in the 63$\,\mu$m line comes from
H$_{2}$-dominated gas and consequently the correlation 
between the surface brightness of this line and the
H$_{2}$ column density is almost non-existent 
(Figure~\ref{fig:LOI_NH2}). For the 145$\,\mu$m line of \oino, 
the situation is very similar: only 30.8\% of the total emission is
produced by H$_{2}$-dominated regions, and there is no 
correlation between the surface brightness of the line and the
H$_{2}$ column density.

\subsubsection{Results from the other simulations}
We have also investigated the behaviour of the \cii and \oi emissivities within our
other three simulations. However, we find that in the Low and Weak simulation,
the \cii and \oi emission is extremely weak, on account of the low gas temperature,
and is unlikely to be detectable with current instrumentation. We therefore do not
examine the distribution of the emission in this simulation in any detail, and focus
our attention on the remaining two simulations.

\begin{figure}
\includegraphics[width=0.5\textwidth]{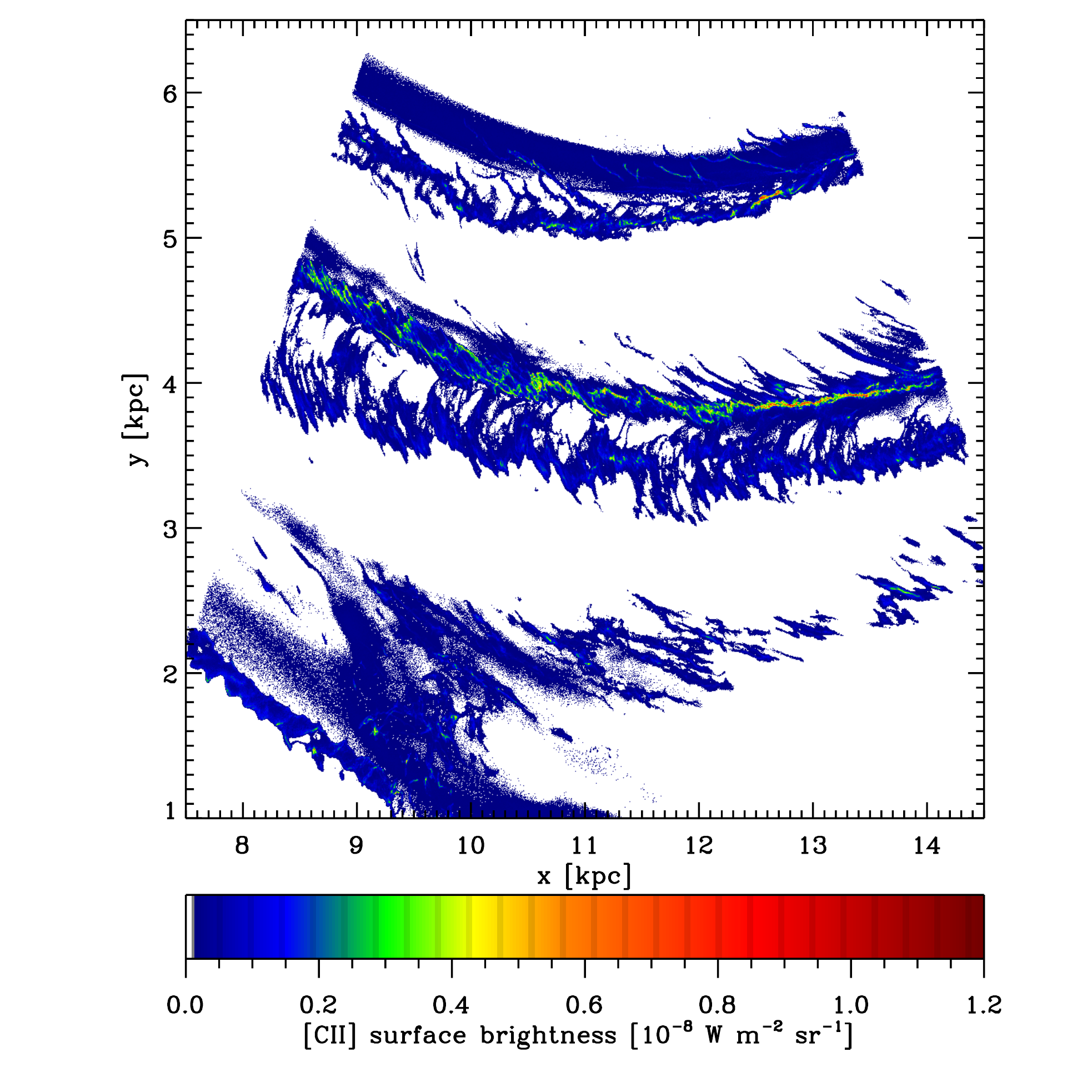}
\caption{As Figure~\ref{fig:CII_SB}, but for the Low Density simulation. \label{fig:CII_SB_LD}}
\end{figure}

\begin{figure}
\includegraphics[width=0.5\textwidth]{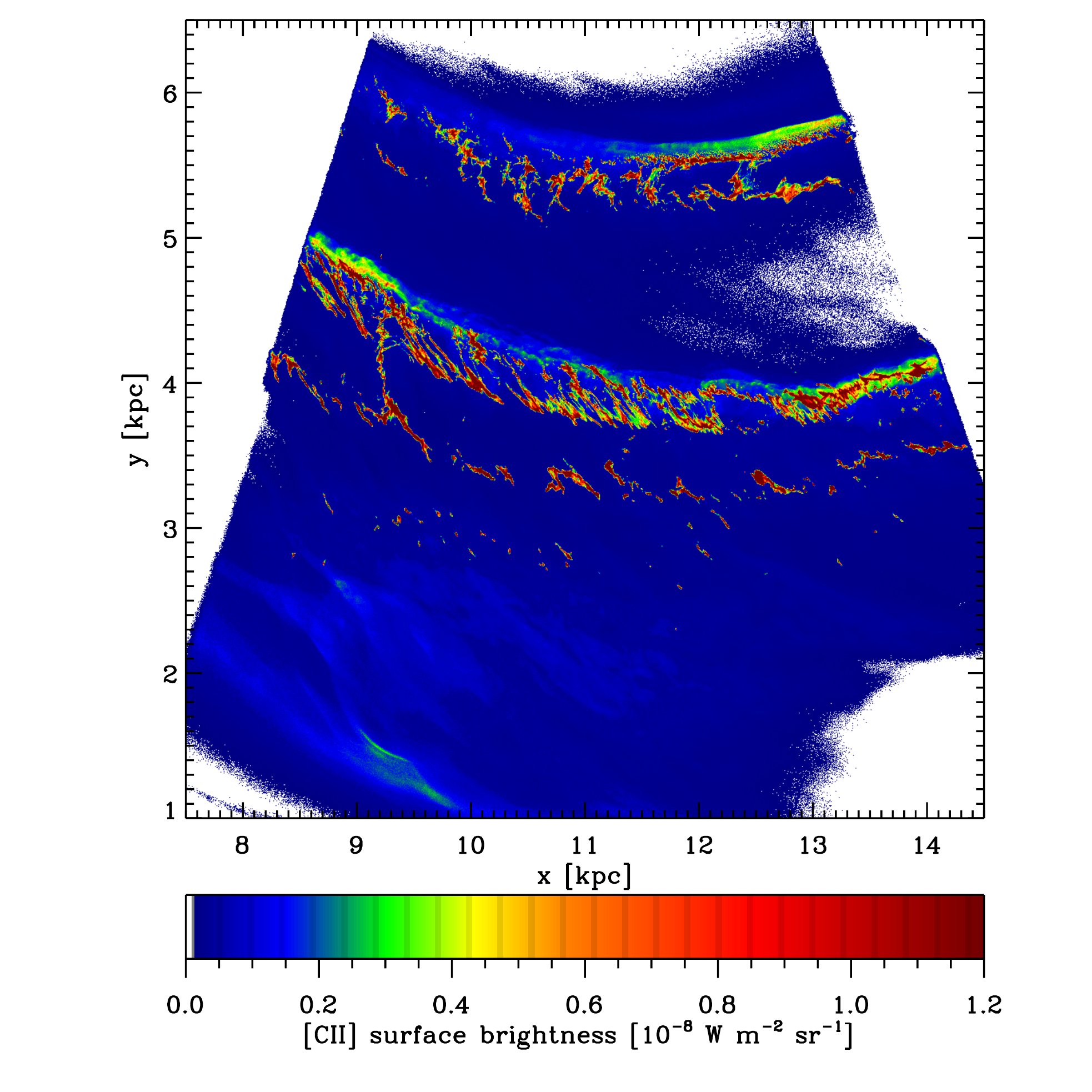}
\includegraphics[width=0.5\textwidth]{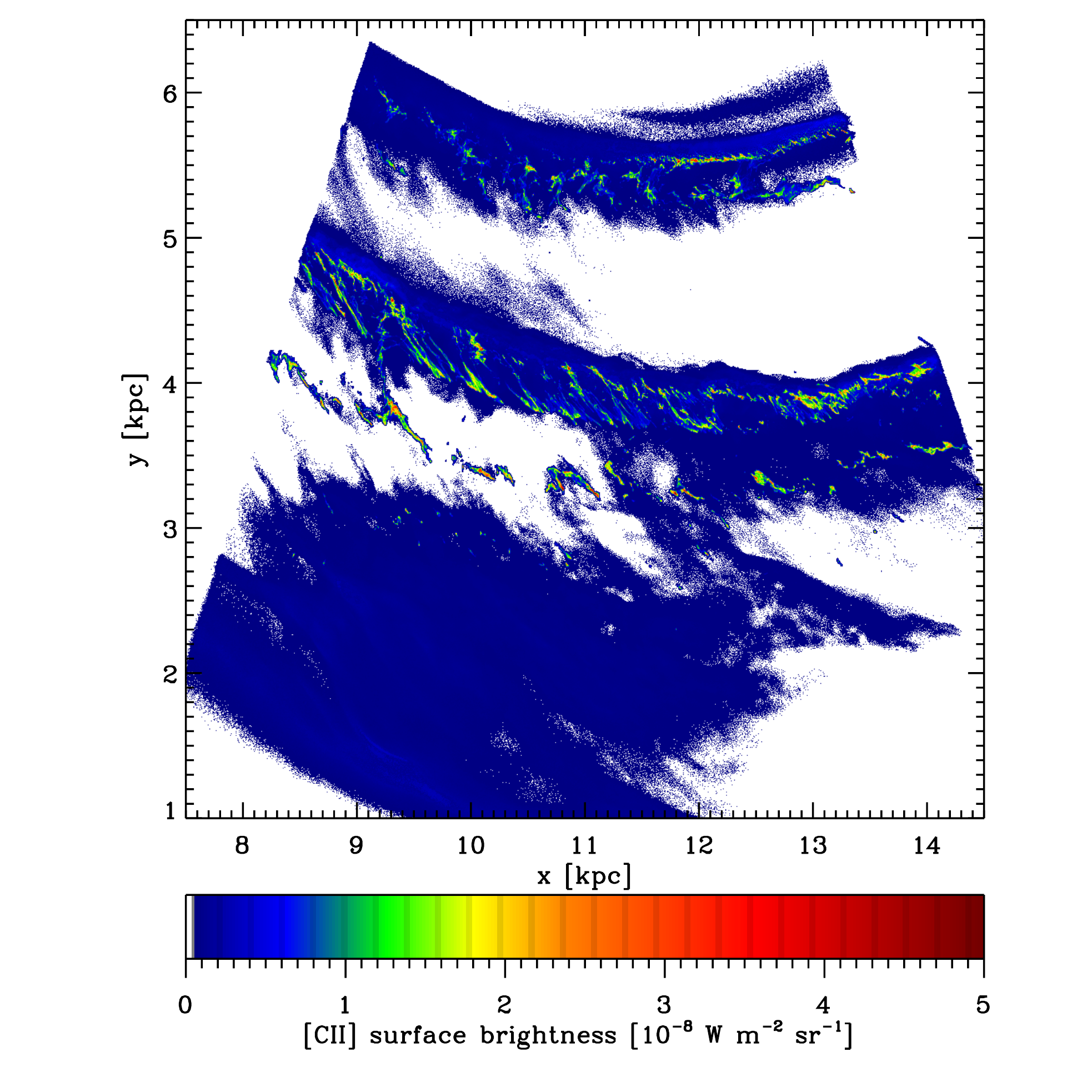}
\caption{{\it Top panel:} As Figure~\ref{fig:CII_SB}, but for the Strong Field simulation. 
The scaling of the colour bar here is the same as in Figure~\ref{fig:CII_SB}, which allows
the emission from the diffuse inter-arm gas to be clearly seen, but leads to saturation in
the brightest regions of the image.
{\it Bottom panel:} A version of the same figure with a rescaled colour bar that better 
represents the behaviour of $I_{\rm CII}$ in the brightest regions.
\label{fig:CII_SB_SF}}
\end{figure}

In Figure~\ref{fig:CII_SB_LD}, we show a map of the \cii surface brightness in our
Low Density simulation at $t = 261.1 \: {\rm Myr}$. Comparing this with Figure~\ref{fig:CII_SB},
we see immediately that although the strength of the \cii emission from the spiral arms is roughly
comparable to that in the spiral arms in the Milky Way simulation, the \cii surface brightness 
from the inter-arm regions is much lower and is unlikely to be detectable. The temperature and
chemical composition of the gas in the inter-arm regions is comparable in both simulations, 
and so the difference in \cii surface brightness is driven primarily by the difference in the surface
density of the gas in these regions in the two simulations. Specifically, we found already in 
\citet{smith14} that much less dense structure formed in the inter-arm regions in the Low Density
simulation than in the Milky Way simulation. This led to a large difference in the CO brightness
of these two regions, and we see here that it also strongly affects the \cii surface brightness of
the inter-arm gas. We see similar behaviour if we look at the spatial distribution of the \cii surface
brightness in the Strong Field simulation (Figure~\ref{fig:CII_SB_SF}). Again, there is less dense structure in the
inter-arm regions than in the Milky Way run, and this is reflected in the \cii emission map. However,
we also see that in the regions where there is significant emission, the \cii surface brightness is
much larger. This is a consequence of the higher temperature of the CO-dark H$_{2}$ and dense
atomic hydrogen in this run, due to the greater photoelectric heating rate. It suggests that it
will be much easier to detect \cii emission from this gas in galaxies that have stronger ISRFs 
than is typical in the local region of the Milky Way.

\begin{figure}
\includegraphics[width=0.5\textwidth]{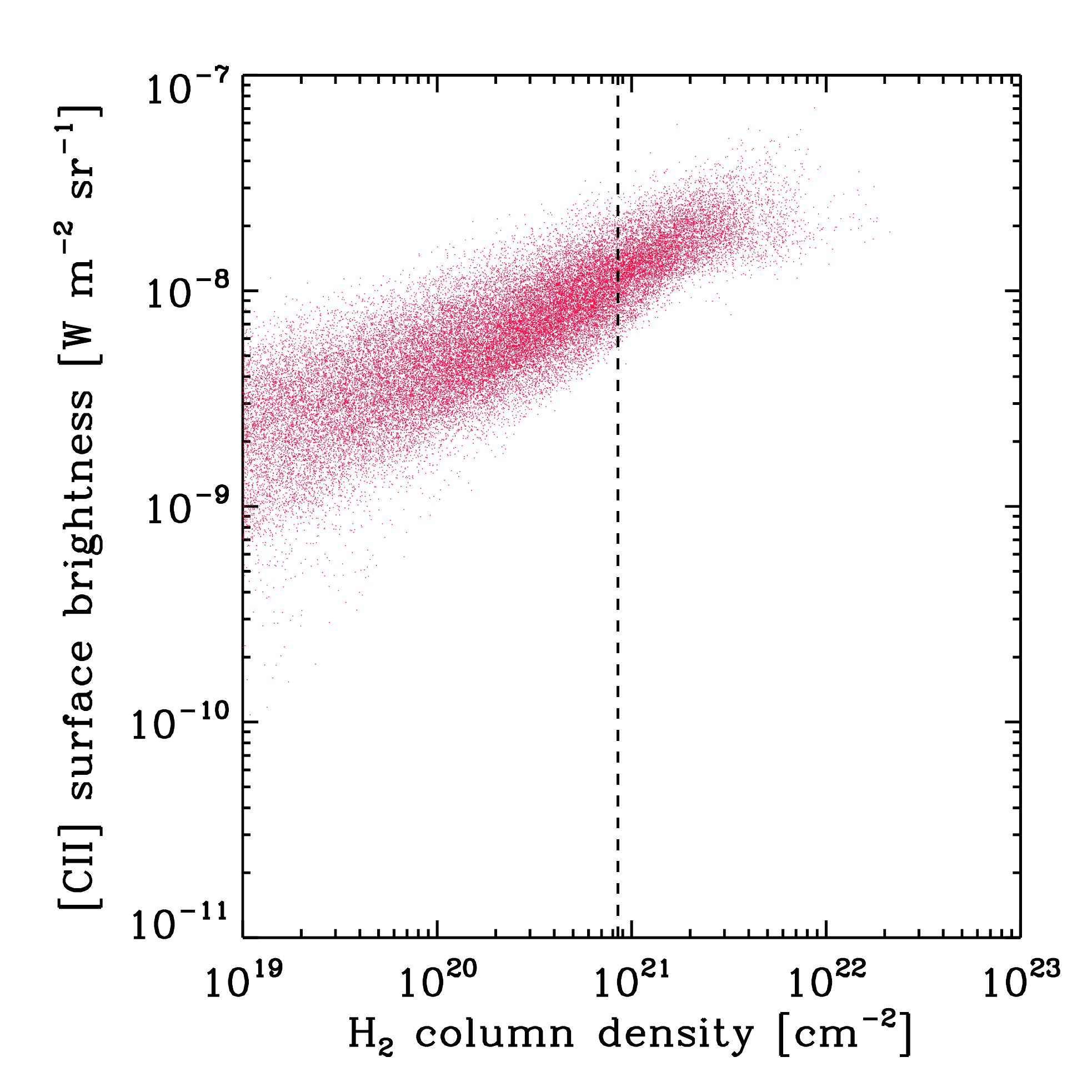}
\caption{As Figure~\ref{fig:LCII_NH2}, but for the Strong Field simulation.
\label{fig:LCII_NH2_SF}}
\end{figure}

The impact of a factor of ten increase in the strength of the ISRF on the \cii surface brightness
becomes even more clear if we examine the correlation between \cii surface brightness and H$_{2}$
column density in the Strong Field run (Figure~\ref{fig:LCII_NH2_SF}). We see that $I_{\rm CII}$ scales 
approximately as $I_{\rm CII} \propto N_{\rm H_{2}}^{1/2}$, just as in the Milky Way run. However, if we compare
this figure with Figure~\ref{fig:LCII_NH2}, we see that at any given H$_{2}$ column density, the \cii surface
brightness in the Strong Field run is a factor of five to ten larger than in the Milky Way run. This behaviour
is easy to understand from an energetics perspective: an increase in the strength of the ISRF by a factor
of ten leads to approximately a factor of ten increase in the photoelectric heating rate. Therefore, in
conditions where \cii dominates the cooling, we should expect the \cii luminosity to also increase by
approximately an order of magnitude. 

\begin{figure}
\includegraphics[width=0.5\textwidth]{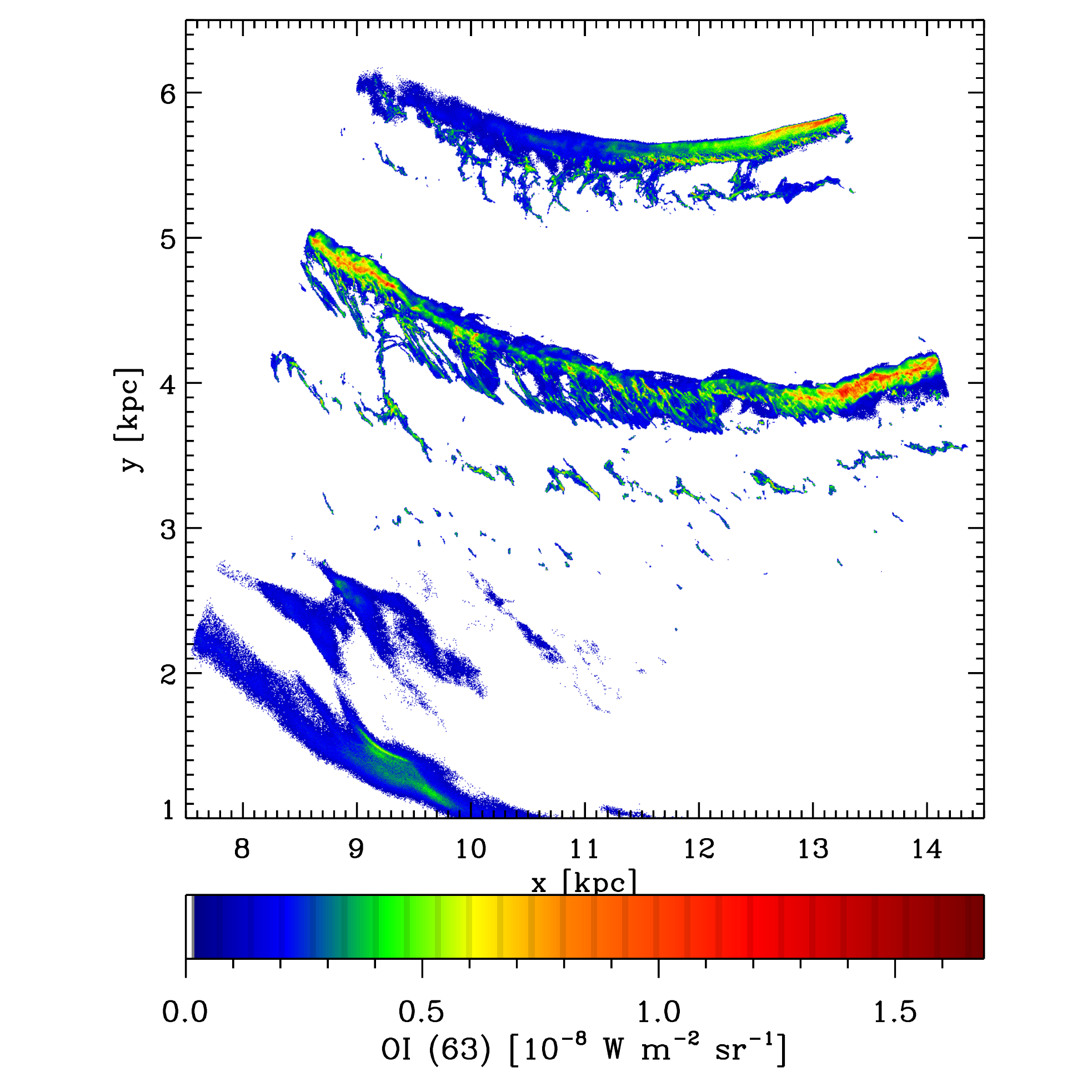}
\caption{Map of the estimated \oi 63~$\mu$m surface brightness in the Strong Field
simulation at time $t = 261.1 \: {\rm Myr}$. Regions shown in white have surface brightnesses $I_{\rm OI} 
\leq 10^{-9} \: {\rm W \, m^{-2} \, sr^{-1}}$.
\label{fig:OI_SB_SF}}
\end{figure}

\begin{figure}
\includegraphics[width=0.5\textwidth]{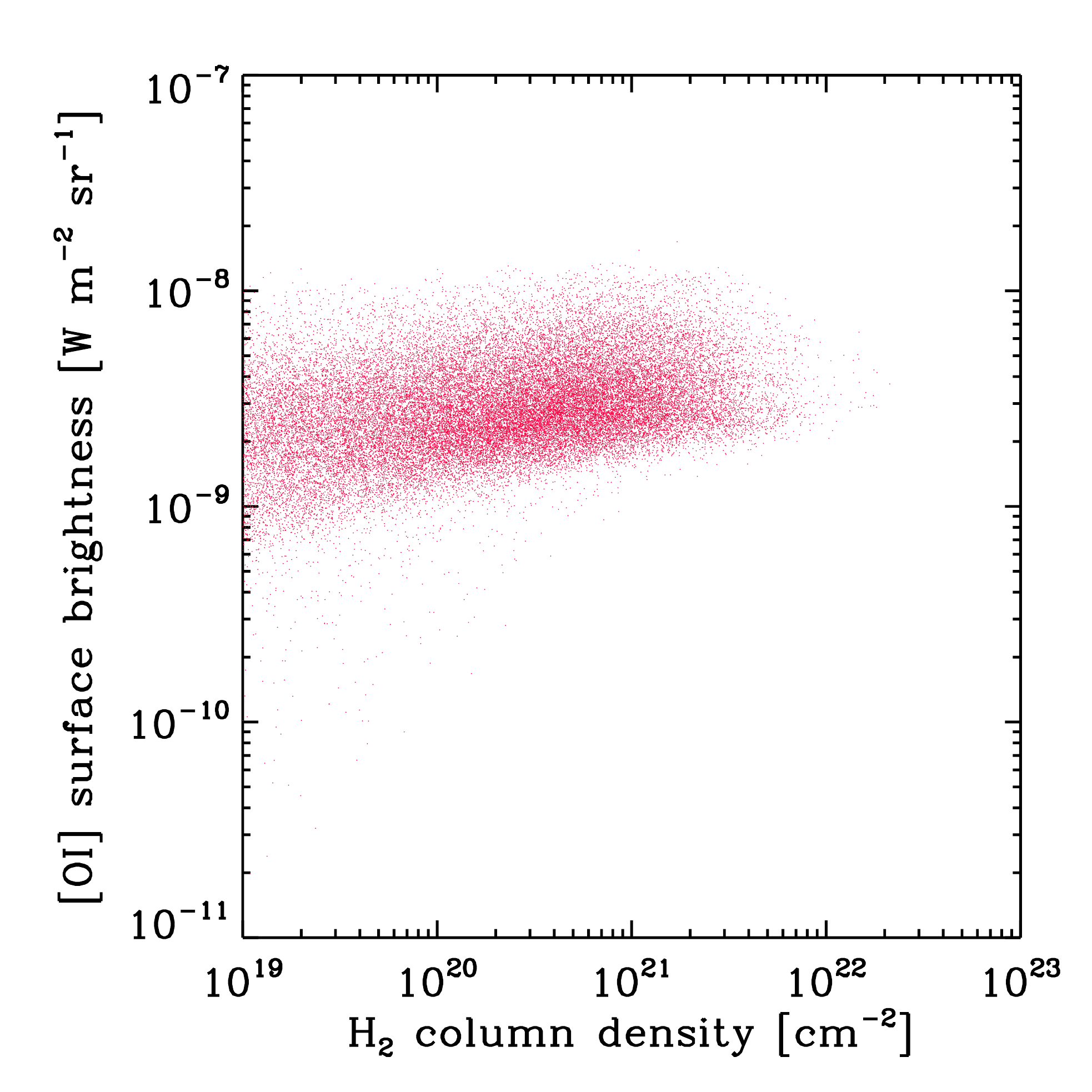}
\caption{Estimated \oi 63~$\mu$m surface brightness, plotted as a function of
the H$_{2}$ column density, for the Strong Field simulation at
time $t = 261.1 \: {\rm Myr}$. \label{fig:OI_H2_SF}}
\end{figure}

Turning to \oino, we start by noting that in the Low Density simulation, the 63~$\mu$m line is relatively weak:
$I_{\rm OI} < 10^{-9} \: {\rm W \: m^{-2} \: sr^{-1}}$ throughout most of the disk, and hence is likely undetectable
with current instruments. The only region in which we find a higher \oi surface brightness is in the dense
spiral arm in the middle of the high-resolution section of the disk. However, even here $I_{\rm OI}$ never exceeds
a few times $10^{-9} \: {\rm W \: m^{-2} \: sr^{-1}}$ and the \oi to \cii ratio, $I_{\rm OI}/I_{\rm CII}$, is much smaller
than one. In the Strong Field simulation, on the other hand, the higher gas temperatures lead to significantly
higher \oi surface brightnesses in the spiral arms, as illustrated in Figure~\ref{fig:OI_SB_SF}.  However, the correlation 
between the \oi surface brightness and the H$_{2}$ column density remains very poor (Figure~\ref{fig:OI_H2_SF}),
and so even in these warmer conditions, \oi emission is not a good tracer of CO-dark H$_{2}$.

\section{Discussion}
\label{sec:discuss}
\subsection{The temperature distribution of CO-dark molecular gas}
Our simulations demonstrate that the temperature distribution of  CO-dark molecular
gas is significantly different from the temperature distribution of the molecular gas 
that is well-traced by CO emission. CO-bright molecular gas has a low mean 
temperature, which can be as low as 10~K or as high as 30~K, depending on the
strength of the ambient ISRF, and also has a narrow temperature distribution.
CO-dark molecular gas, on the other hand, has a much higher mean temperature,
ranging from 74~K in the Low and Weak simulation to 105~K in the Strong Field
simulations. The CO-dark H$_{2}$ also occupies a broader range of temperatures
than the CO-rich gas. It is clear from this that
temperatures derived using molecular tracers (e.g.\ CO itself or NH$_{3}$) are
only useful for describing the behaviour of the CO-bright gas, and not for constraining
the temperature of the CO-faint material.

In addition, if we compare the temperature distribution of the cold H$\,${\sc i} in our
simulations with that of the CO-dark H$_{2}$, we see that although there are more
similarities, the two temperature distributions are not the same. H$\,${\sc i} is typically
found in slightly warmer gas than CO-dark H$_{2}$, since the conditions that favour
H$_{2}$ formation (higher densities, moderate amounts of shielding) also allow the
gas to reach lower temperatures than the lower density gas dominated by atomic
hydrogen. Therefore, temperature measurements made using H$\,${\sc i} emission
or absorption are also only of limited usefulness in constraining the temperature
of the CO-dark H$_{2}$.
  
One consequence of this is that the sound speed of the CO-dark molecular gas is
significantly lower than that of the cold H$\,${\sc i}. If we compute the mass-weighted
mean sound speed of the regions dominated by cold H$\,${\sc i} in our Milky Way
simulation (i.e.\ regions with molecular fractions of less than 50\% and temperatures
$T < 300$~K), then we find that $c_{\rm s, CNM} = 1.15 \: {\rm km \: s^{-1}}$.
This value is only weakly dependent on the value we take for the upper limit on $T$,
provided that it is higher than a few hundred K. For comparison, the mass-weighted mean sound
speed of the H$_{2}$-dominated regions (i.e.\ those with molecular fractions greater than 
50\%) is $c_{\rm s, H_{2}} = 0.47 \: {\rm km \: s^{-1}}$
if we consider all of the H$_{2}$-dominated regions, increasing to 
$c_{\rm s, dark} = 0.64 \: {\rm km \: s^{-1}}$ if we only
include cells in which less than 1\% of the total carbon is locked up in CO.

This difference in sound speeds has two important consequences. First, it implies that
the Jeans mass in the CO-dark molecular gas will generally be smaller than in the 
H$\,${\sc i}-dominated regions, meaning that the molecular regions will be more
unstable to gravitational collapse. Second, if the turbulent velocity dispersion in
the H$\,${\sc i}-dominated gas is similar to that in the CO-dark H$_{2}$, then 
turbulence will have an easier job creating dense structures in the latter than in the
former. For example, we know that for an isothermal, supersonically turbulent gas,
the logarithmic density variance scales as \citep{molina12}
\begin{equation}
\sigma^{2} = \ln \left[1 + b^{2} \left(\frac{\beta}{\beta + 1} \right) {\cal M}^{2} \right], 
\end{equation}
where ${\cal M} \equiv \sigma_{v} / c_{\rm s}$ is the Mach number of the
turbulence, $\beta$ is the ratio of the thermal pressure to the magnetic 
pressure, given by  $\beta \equiv 2 c_{\rm s}^{2} / v_{\rm A}^{2}$ where
$v_{\rm A}$ is the Alfven velocity,  and 
$b$ is a parameter that depends on the nature of the turbulent forcing
\citep{federrath08,federrath10}. In a non-isothermal gas, one recovers a slightly
different relationship between $\sigma$ and ${\cal M}$ \citep{nolan15}, but the
general point that an increase in ${\cal M}$ leads to an increase in the density
variance (i.e.\ an increased number of highly over-dense or under-dense regions)
remains valid.
Together, these factors mean that it is easier to form gravitationally bound
clouds in regions of CO-dark molecular gas than in regions of cold atomic gas.
It also implies that simulations of cloud formation in the ISM that do not account
for the transition from H$\,${\sc i} to H$_{2}$ will tend to underestimate the
formation rate of gravitationally bound clouds.

Another important implication of the fact that the CO-dark H$_{2}$ has a
relatively broad temperature distribution concerns its detectability using the fine structure
lines of \cii and \oino. The excitation rate of \cii due to collisions with H$_{2}$ increases
by a factor of ten as we increase the temperature from 30~K to 100~K, and so the warmer
the H$_{2}$, the easier it is to trace using \cii emission. Consequently, attempts to map
the distribution of CO-dark molecular gas using \cii emission \citep[e.g.][]{langer14}
inevitably give us a somewhat biased picture, as they preferentially pick out warmer gas and may not
have sufficient sensitivity to detect CO-dark H$_{2}$ in relatively cold regions. In the case
of \oino, the excitation rate changes by an even larger factor over the temperature
range $30 < T < 100$~K, and so the bias is even greater.
      
\subsection{Caveats}
\label{sec:caveats}
As previously discussed in \citet{smith14}, there are several caveats that should be taken into account when considering the results of our
simulations. First, and most importantly, there is the fact that in the simulations discussed here, we do not account for the effects 
of either self-gravity or stellar feedback. Although we expect the effects of these processes to cancel out to some extent, with
stellar feedback disrupting clouds that would otherwise undergo runaway gravitational collapse, our neglect of both processes
nonetheless represents a major simplification in comparison to the real ISM. Note, however, that we would expect both processes
to have a much stronger impact on the behaviour of the relatively dense, CO-rich molecular gas than on the comparatively
diffuse CO-dark H$_{2}$.

A second important simplification in our models is our assumption of a uniform interstellar radiation field. On small scales
(e.g.\ close to regions of massive star formation), this assumption clearly breaks down, but on large scales it should be
a reasonable approximation \citep{Wolfire03}. 
Our assumption of a constant metallicity and a constant dust-to-gas ratio also represents a simplification in comparison
to the real Milky Way, which has a slight metallicity gradient \citep[see e.g.][]{Luck11}, although for the range of radii we
simulate, this corresponds to a change in metallicity of less than a factor of two. 

Finally, the treatment of CO chemistry used in these simulations (based on \citealt{nl97}) is somewhat approximate and
is known to over-produce CO in turbulent clouds in comparison to more sophisticated models \citep{gc12a}. Consequently,
we may mis-classify as CO-rich some gas that should actually be CO-dark. However, accounting for this will only broaden
the temperature distribution of the CO-dark H$_{2}$, strengthening our main conclusions.

\section{Summary}
\label{sec:conc}
We have investigated the temperature distribution of molecular gas in a set of simulations of a representative portion of
the ISM of a disk galaxy. Our simulations were carried out using the {\sc Arepo} moving-mesh code and include a treatment
of the non-equilibrium chemistry and thermal evolution of the gas in the ISM. For simplicity, in our current simulations we
do not account for the effects of self-gravity, star formation or stellar feedback, but we do account for the influence of the 
large-scale interstellar radiation field.

In our fiducial run, which has a surface density and ISRF strength characteristic of the local ISM, we find that there is a
clear difference between the temperature distribution of CO-bright and CO-dark molecular gas. CO-bright regions have
a narrow distribution of temperatures, $10 < T < 30$~K, in good agreement with observational estimates of temperatures
in Galactic GMCs. However, the temperature distribution of the CO-dark gas is much broader, covering the range $10 <
T < 100$~K. This temperature range overlaps with the range of temperatures that we find in regions dominated by cold
atomic hydrogen, but the two temperature distributions are not the same: the CO-dark molecular gas is colder, on average,
than the atomic gas. This implies that the temperatures that we can infer from H$\,${\sc i} observations of cold atomic gas
are not a reliably proxy for the temperatures in the CO-dark gas. An important consequence of this is the fact that the mean
sound speed of the CO-dark molecular component is significantly smaller than that of the cold atomic gas, differing by almost 
a factor of two in our fiducial run. As a result, the molecular component will be more susceptible to turbulent compression
and gravitational collapse than the atomic gas, implying that the chemical transition from H to H$_{2}$ can help to promote
star formation even though the H$_{2}$ molecules themselves play no role in cooling the gas at these temperatures \citep{gc12c}.

We have explored how these results vary as we vary the initial surface density of the gas and the strength of the ISRF. 
Decreasing the surface density while keeping the ISRF strength fixed leads to a large change in the temperature
distribution of the molecular gas, owing to the substantially decreased effectiveness of shielding in this case. The
CO-bright gas almost entirely vanishes \citep[c.f.][]{smith14} and the temperature distribution of the H$_{2}$ changes
from being relatively flat between 10~K and 100~K to being sharply peaked around 100~K. On the other hand, if we
decrease the strength of the ISRF at the same time as we decrease the surface density, we find that we retain the
cold H$_{2}$. Indeed, in this case, there is significantly more H$_{2}$ with $T \sim 20$--30~K and significantly less
with $T \sim 100$~K than in our fiducial run, despite the lower surface density. Finally, if we keep the surface density
of the gas fixed and increase the strength of the ISRF, we find that the temperature distribution of the H$_{2}$ again
becomes strongly peaked around $T \sim 100$~K. However, in this case, unlike in the low surface density run, we
retain a significant amount of relatively cold H$_{2}$. 

We have also investigated the potential detectability of the CO-dark H$_{2}$ using the [C$\,${\sc ii}] and [O$\,${\sc i}] 
fine structure lines. The wide range of spatial scales covered by our simulations means that it is impractical to 
post-process them using a radiative transfer code in order to compute the integrated intensities of the lines without
substantial smoothing. We have therefore estimated the line strengths using a simple approach that does not require
us to smooth the data. The main approximation made in this approach is the assumption that the lines remain 
optically thin, which is valid for low column density lines of sight but causes us to overestimate the emission for
lines of sight with $A_{\rm V} > 1$.

In our fiducial simulation, we find that roughly equal amount of emission are produced in regions dominated by
atomic hydrogen and in regions dominated by CO-dark H$_{2}$. In the latter regions, the 
typical [C$\,${\sc ii}] surface brightness ranges from a few times $10^{-9} \: {\rm W \: m^{-2} \: sr^{-1}}$ up to
$10^{-8} \: {\rm W \: m^{-2} \: sr^{-1}}$. Although faint, this is in principle detectable with modern instruments
such as FIFI-LS, albeit with relatively low signal-to-noise. However, even if detected, interpreting this emission
may be difficult: distinguishing between emission from regions dominated by atomic hydrogen and those
dominated by CO-dark H$_{2}$ may not be straightforward
and there is only a weak correlation between the [C$\,${\sc ii}] surface brightness and the H$_{2}$
column density. For the [O$\,${\sc i}] lines, the situation is less promising: both lines are significantly fainter
and there is essentially no correlation between their surface brightness and the H$_{2}$ column density.

Comparing the results from our fiducial simulation to those from our other simulations, we find that the most
important parameter controlling the detectability of the CO-dark molecular gas with the \cii and \oi lines is the strength
of the ISRF. Increasing the strength of the ISRF increases the photoelectric heating rate and hence the temperature
of the CO-dark molecular gas, causing it to emit more strongly in \cii and \oino. Conversely, decreasing the strength of 
the ISRF makes this gas colder and hence much harder to map using these lines. We also find that even in
the most favourable conditions for detecting \cii and \oino, their usefulness as a tracer of CO-dark H$_{2}$ is
questionable: \cii emission is only weakly correlated with H$_{2}$ column density, and \oi emission is almost
completely uncorrelated with H$_2$.

\section*{Acknowledgements}
The authors would like to thank P.~Clark, R.~S.~Klessen, J.~Pineda and V.~Springel for useful discussions. They would also like to thank the anonymous
referee for their comments on a earlier draft of this paper.
SCOG acknowledges financial support from the Deutsche Forschungsgemeinschaft (DFG) via SFB 881 ``The Milky Way System" (subprojects B1, B2 and B8)
and via SPP 1573 ``Physics of the Interstellar Medium'' (grant number GL 668/2-1). RJS gratefully acknowledges support from the Royal Astronomical Society
through their Norman Lockyer Fellowship. The numerical simulations described in this paper were performed on the Milky Way supercomputer, which is funded 
by the DFG through SFB 881 (subproject Z2) and hosted and co-funded by the J\"ulich Supercomputing  Center (JSC).

\appendix
\section{When do the \cii and \oi lines become optically thick?}
\label{app:opthick}
We can assess the conditions in which the optically thin approximation
for treating \cii and \oi emission is valid
by computing the column density of ground state C$^{+}$ ions
or oxygen atoms needed to produce an optical depth $\tau = 1$ in
the 158$\,\mu$m or 63$\,\mu$m lines, respectively. For the purposes
of this calculation, we conservatively assume that the broadening of
both lines is dominated by turbulent motions with a one-dimensional
velocity dispersion of $\sigma_{\rm 1D} = 1 \: {\rm km \: s^{-1}}$. In practice, 
it is likely that in many of the CO-dark regions, $\sigma_{\rm 1D}$ will be
higher than this, meaning that the required column densities will be larger.

We find that with this assumption, the \cii 158$\,\mu$m line becomes optically
thick once the C$^{+}$ column density exceeds $N_{\rm C^{+}} \simeq
2.4 \times 10^{17} \: {\rm cm^{-2}}$. If all of the carbon along the line-of-sight
is in the form of C$^{+}$, then this corresponds to a total hydrogen column
density $N_{\rm H, tot} = N_{\rm C^{+}} / x_{\rm C, tot} \simeq 1.7 \times 10^{21}
\: {\rm cm^{-2}}$, or a visual extinction of $A_{\rm V} \sim 1$. This is close to
the point at which carbon starts to become locked up in CO in our simulations,
and implies that by assuming optically thin emission, we may overestimate the
emission coming from ionized carbon gas in the immediate vicinity of GMCs,
but will probably not significantly overestimate the emission coming from more
diffuse regions. For the \oi 63$\,\mu$m line, the column density of atomic
oxygen required to produce an optically thick line is somewhat larger,
$N_{\rm O} \simeq 6 \times 10^{17} \: {\rm cm^{-2}}$. However, the total
oxygen abundance in the ISM is also larger, and so we expect the line to
become optically thick at roughly the same hydrogen column density.
Therefore, the same caveats apply as in the case of \ciino. 

Note also that in both cases, if the emission is produced primarily in gas with
a density much less than the critical density of the transition, then the line
can be effectively thin even if $\tau > 1$. In this regime, discussed in more
detail in \citet{gold12} for the case of [C$\,${\sc ii}], the photons are absorbed and
re-emitted multiple times, but the fraction of the radiated energy that is
converted back into heat by collisional de-excitation of excited C$^{+}$ or
O remains very small.

Finally, we note that as the density of the gas in our simulations is everywhere much 
smaller than the critical density of \oino, $n_{\rm crit} \simeq 10^{6} \: {\rm cm^{-3}}$, 
we would not expect the \oi 145$\,\mu$m line to be optically thick anywhere 
within our simulations.

\end{document}